\documentclass[useAMS,usenatbib]{mnras}

\usepackage{color}
\usepackage{threeparttable}         %to use 
\usepackage{graphicx,epsfig}
\usepackage{multicol}
\usepackage{hyperref}
\usepackage{amsmath,amssymb}
\usepackage[T1]{fontenc}
\usepackage{aecompl}

\def\aj{{AJ}}                   % Astronomical Journal
\def\apj{{ApJ}}                 % Astrophysical Journal
\def\apjl{{ApJ}}                % Astrophysical Journal, Letters
\def\apjs{{ApJS}}               % Astrophysical Journal, Supplement
\def\ao{{Appl.~Opt.}}           % Applied Optics
\def\aap{{A\&A}}                % Astronomy and Astrophysics
\def\aaps{{A\&AS}}              % Astronomy and Astrophysics, Supplement
\def\mnras{{MNRAS}}             % Monthly Notices of the RAS
\def\pasp{{PASP}}               % Publications of the ASP
\def\pasj{{PASJ}}               % Publications of the ASJ
\def\nat{{Nature}}              % Nature
\def \hmpc{~h^{-1}\,{\rm Mpc}}

\def \gsim { \lower .75ex \hbox{$\sim$} \llap{\raise .27ex \hbox{$>$}} }
\def \lsim { \lower .75ex \hbox{$\sim$} \llap{\raise .27ex \hbox{$<$}} }

\def\be{\begin{equation}}
\def\ee{\end{equation}}

\def\hmodot{~h^{-1}\rm{M_\odot}}

\renewcommand\TPTminimum{8.5cm}
\title[The VLT LBG Redshift Survey V]{The VLT LBG Redshift Survey - V. Characterising the $z=3.1$ Lyman Alpha Emitter Population}
\author[R. M. Bielby et al.]{R. M. Bielby$^{1}$, P. Tummuangpak$^{1,2}$\thanks{pimptu@kku.ac.th}, T. Shanks$^{1}$, H. Francke$^{3}$, N. H. M. Crighton$^{4}$,
\newauthor E. Ba\~{n}ados$^{5,6}$, Jorge Gonz\'alez-L\'opez$^{5}$, L. Infante$^{5}$, A. Orsi$^{7}$ \\
$^{1}$Department of Physics, University of Durham, South Road, Durham DH1 3LE, UK\\
$^{2}$Department of Physics, Khon Kaen University, Muang, Khon Kaen, 40002, Thailand\\
$^{3}$Joint ALMA Observatory/European Southern Observatory, Alonso de Cordova 3107, Vitacura, Santiago, Chile\\
$^{4}$Centre for Astrophysics and Supercomputing, Swinburne University of Technology, PO Box 218, Victoria 3122, Australia\\
$^{5}$Institute of Astrophysics, Center for Astro-Engineering, Pontificia Universidad Cat\'{o}lica de Chile, Av. Vicu\~{n}a Mackenna 4860, \\
7820436 Macul, Santiago, Chile\\
$^{6}$Max Planck Institut f\"ur Astronomie, K\"onigstuhl 17, D-69117, Heidelberg, Germany\\
$^{7}$Centro de Estudios de F\'isica del Cosmos de Arag\'on, Plaza San Juan 1, Planta-2, E-44001 Teruel, Spain
}

\begin{document}
\date{Accepted 2015 December 10. Received 2015 December 10; in original form 2014 December 15}
%\date{}
\volume{456}
%\pagerange{\pageref{firstpage}--\pageref{lastpage}} \pubyear{2016}
\pagerange{4061--4080} \pubyear{2016}

\maketitle
\label{firstpage}

\begin{abstract}
We present a survey of $z\sim3$ Ly$\alpha$ emitters (LAEs) within the fields of the VLT LBG Redshift Survey. The data encompasses 5 independent survey fields co-spatial with spectroscopic LBG data and covering a larger total area than previously analysed for LAE number counts and clustering. This affords an improved analysis over previous work by minimising the effects of cosmic variance and allowing the cross-clustering analysis of LAEs and LBGs. Our photometric sample consists of $\approx600$ LAE candidates, over an area of 1.07~deg$^2$, with equivalent widths of $\gtrsim65$~\AA\ and a flux limit of $\approx2\times10^{-17}$~erg~cm$^{-2}$~s$^{-1}$. From spectroscopic follow-up, we measured a success rate of $78\pm18\%$. We find the $R$-band continuum luminosity function to be $\sim10\times$ lower than the luminosity function of LBGs at this redshift, consistent with previous studies. Exploiting the large area of the survey, we estimate the LAE auto-correlation function and find a clustering length of $r_0=2.86\pm0.33~h^{-1}$~Mpc, low compared to the $z\sim3$ LBG population, but somewhat higher than previous LAE measurements. This corresponds to a median halo mass of $M_{\rm DM}=10^{11.0\pm0.3}~h^{-1}~$M$_{\odot}$. We present an analysis of clustering length versus continuum magnitude and find that the measurements for LAEs and LBGs are consistent at faint magnitudes. Our combined dataset of LAEs and LBGs allows us to measure, for the first time, the LBG-LAE cross-correlation, finding a clustering length of $r_0=3.29\pm0.57~h^{-1}$~Mpc and a LAE halo mass of $10^{11.1\pm0.4}~h^{-1}$~M$_{\odot}$. Overall, we conclude that LAEs inhabit primarily low mass halos, but form a relatively small proportion of the galaxy population found in such halos.
\end{abstract}


\section{Introduction}
\label{sec:intro}

Two key methods for efficiently selecting high redshift galaxies are via their Lyman break feature (Lyman Break Galaxies - LBGs) and via their Lyman $\alpha$ emission (Lyman-$\alpha$ emitters - LAEs). Selecting high redshift galaxies through their strong emission in the Ly$\alpha$ feature using the narrow-band imaging method has come to be a very effective technique to isolate high redshift galaxies. There are many observations that have been made to uncover galaxies with strong Ly$\alpha$ emission at various redshifts \citep[e.g.][]{Hu1996,Hu1998,1998AJ....115.1319C,steidel96,2003ApJ...582...60O,Hayashino2004,gawiser07,Gronwall2007,2008ApJS..176..301O,2008ApJ...681..856R,2010ApJ...723..869O,2013ApJ...773..178B}.

At redshift $z \sim 3$, there are a number of photometric and spectroscopic LAE samples \citep[e.g.][]{Steidel2000, Fynbo2003, Hayashino2004, Matsuda2005, Venemans2007, Nilsson2007, gawiser07, Gronwall2007, 2008ApJ...681..856R, 2008ApJS..176..301O, 2014MNRAS.439..446M}. \citet{Gronwall2007} surveyed the Extended Chandra Deep Field-South (ECDFS), identifying $\approx160$ photometrically selected $z=3.1$ LAEs over an area of 0.28~deg$^2$ and to a Ly$\alpha$ luminosity limit of $L_{\rm Ly\alpha}\gtrsim10^{42}~{\rm erg~s}^{-1}$. Their colour constraints correspond to a \emph{rest-frame} equivalent width limit of ${\rm EW}\gtrsim20$~\AA. By measuring the Ly$\alpha$ and rest-frame UV continuum luminosity functions in combination with the equivalent width distribution, they surmised that LAEs contain a non-negligible amount of dust and are therefore not of a primordial origin (but that they do represent a young galaxy population).  \citet{2008ApJS..176..301O} presented a survey of $\approx350$ photometrically identified (with 41 spectroscopically confirmed) LAEs over $\approx1~{\rm deg}^2$ in the Subaru/XMM-Newton Deep Survey (SXDS)~field to a limiting Ly$\alpha$ luminosity of $L_{\rm Ly\alpha}\gtrsim10^{42}~{\rm erg~s}^{-1}$ (and a rest-frame equivalent width limit of $\gtrsim65$~\AA). Combining their $z=3.1$ data with higher redshift samples, they find that LAEs (at a given equivalent width and luminosity limit) are more common at earlier epochs and are a tracer of either low-extinction (when compared to LBGs) or younger stellar populations.

Although LAEs in general have been widely studied, their clustering properties remain comparatively sparsely studied, due to the need for large survey fields. At $z\approx3$, \citet{Hayashino2004}, \citet{gawiser07} and \citet{2010ApJ...723..869O} have each presented measurements of the LAE angular correlation function. \citet{Hayashino2004} do not present a quantification of their clustering result, whilst \citet{gawiser07} and \citet{2010ApJ...723..869O} measure clustering lengths of $r_0=2.52^{+0.6}_{-0.7}~\hmpc$ and $r_0=1.70^{+0.39}_{-0.46}~\hmpc$ respectively. These two measurements are based on differing equivalent width constraints of ${\rm EW}>20$~\AA\ and ${\rm EW}>65$~\AA\ respectively, with the lower EW cut giving a larger clustering length (although the two measurements only differ by $\approx1\sigma$). At higher redshifts, LAE clustering measurements have been obtained at $z\sim4$ \citep{kovac07,2010ApJ...723..869O}, $z\sim5$ \citep{2003ApJ...582...60O,2004ApJ...605L..93S,2009ApJ...696..546S}, and $z\sim6-7$ \citep{2010ApJ...723..869O}. In each case, these clustering measurements are based on single observational fields of at most $1~{\rm deg}^2$, whilst in some cases \citep[e.g.][ at $z=3.1$]{Hayashino2004} they are actively targeted on high density regions. Indeed, cosmic variance can be a significant source of error in LAE clustering studies \citep{2004ApJ...605L..93S,2009NewAR..53...47F} and measurements based on single fields can prove unrepresentative of the population as a whole. The largest of these works thus far remains that of \citet{2010ApJ...723..869O}, which presented LAE clustering results at $z=3.1$, $z=3.7$, $z=5.7$, and $z=6.6$. An apparent evolution in the LAE clustering was observed, with $r_0$ (and therefore the clustering bias) increasing with increasing redshifts, implying a constant host halo mass with redshift of $\approx10^{11}~h^{-1}{\rm Mpc}$. 

% In terms of understanding the LAE population through galaxy formation models, a number of groups have presented analyses based on semi-analytical models \citep[e.g.][]{2005MNRAS.357L..11L,2007ApJ...670..919K,2008MNRAS.391.1589O,2010ApJ...708.1119K,Orsi2012}. For example, both \citet{2010ApJ...708.1119K} and \citet{Orsi2012} presented results from semi-analytical models, fitting to the observed Ly$\alpha$ luminosity function, making predictions for the nature of LAEs and their interstellar medium (ISM). Based on their Ly$\alpha$ equivalent widths and the continuum luminosity function results, \citet{2010ApJ...708.1119K} focus on the nature of the ISM, suggesting that a clumpy dust distribution might explain the observed high equivalent widths. \citet{Orsi2012} meanwhile implemented a full radiative transfer treatment in combination with the GALFORM model and focussed on the ability of galactic outflows to reproduce the observed LAE properties. They concluded that a shell outflow model with constant wind velocity reproduces well the observed LAE Ly$\alpha$ and continuum luminosity functions and that LAEs have lower metallicities, lower star formation rates and larger sizes than the bulk of the high redshift galaxy population. Additionally, in an earlier paper, \citet{2008MNRAS.391.1589O} studied the clustering of simulated LAEs also using the GALFORM model, finding little dependence of the clustering measurements (i.e. $r_0$, clustering bias and median halo mass) on the Ly$\alpha$ luminosity of LAEs, but a strong dependence on redshift. 

Here we present observations of LAEs in the deep imaging fields of the VLT LBG Redshift Survey (VLRS) aimed at studying the clustering properties of LAE galaxies. Our LAE sample extends over 5 independent observational fields each measuring $\sim0.5^{\circ}\times0.5^{\circ}$ in area, thus combatting the effects of cosmic variance. These fields were selected based on the presence of bright quasars at $z\gtrsim3$ and so are not specifically targeted at galaxy overdensities as with some previous works \citep[e.g.][]{Hayashino2004}. We target the $z\approx3$ LAE population in order to overlap with our $z\approx3$ spectroscopic LBG sample in these fields, allowing cross analysis between the two populations. The primary aim of the VLRS has been to probe the intergalactic medium (IGM) of $z\sim3$ galaxies by observing galaxies close QSO sightlines \citep{2011MNRAS.414...28C,2014MNRAS.442.2094T}. At present the survey consists of 9 fields in total, with spectroscopic observations of Lyman break galaxies (LBGs) taken with VLT VIMOS. The primary survey is described in detail by \citet{2011MNRAS.414....2B,2013MNRAS.430..425B}. By adding narrow-band data to our survey data, we aim to enhance the scope of the survey, probing the volume for both LAEs and other sources of Ly$\alpha$ photons (e.g. Ly$\alpha$ blobs and Ly$\alpha$ emission from and around QSOs).
	
In this paper, we present the narrow band data and analysis of LAE galaxies within the VLRS. The data covers 5 independent survey fields, giving the significant advantage of reducing the potential impact of cosmic variance on our results compared to past LAE surveys. We discuss our deep imaging data in these fields and the selection of LAEs at $z\approx3.1$ using the NB497 narrow band filter in Sec.~\ref{sec:imaging}. We then report on spectroscopic follow up observations and our optimised selection criteria based on these in Sec.~\ref{sec:vimos}. In Sec.~\ref{sec:lumfunc}, we present narrow band number counts and the rest-frame UV continuum luminosity function for our LAE sample, whilst in Sec.~\ref{sec:clustering} we present an analysis of the clustering of the LAEs. Finally we present a discussion of our results in Sec.~\ref{sec:discussion} and our conclusions in Sec.~\ref{sec:conclusions}.

Throughout this work, we adopt a cosmology given by $(\Omega_m, \Omega_\Lambda, h, \sigma_8) = (0.25, 0.75, 0.73, 0.83)$, where h is the Hubble constant in units of 100~km/s/Mpc. Distances and volumes are given in comoving coordinates unless otherwise stated. All magnitudes are given in the AB system unless otherwise stated.

\section{Imaging data}
\label{sec:imaging}

The VLRS consists of 9 deep fields (each measuring between 0.25~deg$^2$ and 1~deg$^2$) centred on bright $z\gtrsim3$ QSOs and containing both deep broad band photometry and spectroscopy \citep{2011MNRAS.414....2B,2013MNRAS.430..425B}. Here we have added to these data by using the Subaru Suprime-Cam instrument \citep{Miyazaki2002} to obtain deep narrow band (497 nm) imaging in 5 of the VLRS fields, which allows for identification of LAEs at $z\sim3.1$ (details in Sec.~\ref{sec:initsel}). The observations were taken 19th September 2009.

Suprime-Cam is a mosaic CCD camera with ten CCDs, measuring $2048\times4096$ pixels, which covers a $34'\times27'$ field of view with a pixel scale of $0.20''$. We observed $\sim0.5^\circ\times0.5^\circ$ fields centred on the bright quasars QSO~B2359$+$068, LBQS~0301$-$0035, QSO~J0124$+$0044, PKS~2126$-$158, and LBQS~2231$-$0015 (details are in Table~\ref{tab:NB-Imaging}). These fields were observed with the narrow band [OIII] NB497 filter \citep[4977~\AA, FWHM~77~\AA; ][]{Hayashino2003}. Individual exposure times were 1,200~s for each frame and the seeing was generally sub-arcsecond ($\sim0.6''-0.7''$). For the purposes of photometric calibration, we observed the standard star LTT~9491 (R.A. 23:19:34.98, Dec. -17:05:29.8 J2000).

\begin{table*}
\centering
\caption{Details of narrowband  imaging (NB497) from the Subaru/Suprime-Cam.}
\begin{tabular}{lccccccccc}
\hline
Central QSO      &  R.A.        & Dec.         & QSO redshift & $T_{\rm exp}$ & Seeing   & $M_{\rm ZP}$ & $1\sigma$ depth \\ 
                 &  \multicolumn{2}{c}{(J2000)}&              & (s)           & ($''$) &              &                 \\
\hline
QSO B2359$+$068  &  00:01:40.57 & +07:09:54.1  & 3.234        & 6000          & 0.61     & 31.94        & 28.38           \\
QSO J0124$+$0044 &  01:24:03.78 & +00:44:32.7  & 3.807        & 4800          & 0.55     & 31.93        & 27.47           \\
LBQS 0301$-$0035 &  03:03:41.05 & -00:23:21.9  & 3.175        & 7200          & 0.60     & 31.93        & 28.57           \\
PKS 2126$-$158   &  21:29:12.18 & -15:38:41.0  & 3.268        & 6000          & 0.96     & 31.89        & 27.90           \\
QSO B2231$-$0015 &  22:34:08.99 & +00:00:01.7  & 3.027        & 6000          & 0.87     & 31.93        & 27.77           \\
\hline
\end{tabular}
\label{tab:NB-Imaging}
\end{table*}	

In order to select LAE galaxies, we also require $B$ and $V$ or $R$ deep broad band imaging, which are available from our previous work in these fields. The full details of the broad band data and its reduction are provided by \citet{bouche04,2011MNRAS.414....2B,2013MNRAS.430..425B}, whilst a summary of the broad band data is given in Table~\ref{tab:BB-Imaging}.

\begin{table*}
\centering
\caption[Details of broadband imaging observations]{Details of broadband imaging observations from \citet{2011MNRAS.414....2B,2013MNRAS.430..425B}.}
\begin{tabular}{lccccl}
\\
\hline
Central QSO      & Band & $T_{\rm exp}$ & Seeing   & $1\sigma$ depth & Instrument  \\
                 &      & (ks)          & (arcsec) &                 &             \\
\hline
QSO B2359$+$068  & $B$  & 7.2           &  1.45    & 28.23           & KPNO/MOSAIC \\
                 & $R$  & 6.0           &  1.15    & 27.76           &             \\
\hline
QSO J0124$+$0044 & $B$  & 2.8           & 1.50     & 27.99           & KPNO/MOSAIC \\
                 & $V$  & 3.1           & 1.40     & 27.74           &             \\
\hline
LBQS 0301$-$0035 & $B$  & 6.4           &  1.28    & 28.30           & KPNO/MOSAIC \\
                 & $R$  & 4.8           &  1.19    & 27.66           &             \\
\hline
PKS 2126$-$158   & $B$  & 7.8           & 1.60     & 28.44           & CTIO/MOSAIC2\\
                 & $R$  & 6.4           & 1.50     & 28.27           &             \\
\hline
QSO B2231$-$0015 & $B$  & 13.2          &  1.01    & 27.28           & WFCam (INT) \\
                 & $R$  & 19.2          &  1.01    & 26.98           &             \\

\hline
\end{tabular}
\label{tab:BB-Imaging}
\end{table*} 

\subsection{Data Reduction}

To reduce the Suprime-Cam raw narrowband data, we used the pipeline software, {\sc sdfred} (the Suprime-Cam Deep field REDuction package\footnote{http://www.naoj.org/Observing/DataReduction/}, \citealt{Ouchi2004-SD}) which  comprises {\sc iraf}, {\sc SExtractor} \citep{sextractor}, and the mosaic-CCD data reduction software \citep{Yagi2002}. The package includes bias subtraction, flat fielding, distortion$+$atmospheric dispersion corrections, matching the PSF size, sky subtraction, masking vignetting caused by the auto-guider, masking bad pixels, image alignments and scaling, and mosaicing. The reduction procedure is briefly described below.

The images were first bias subtracted using the median value of the overscan region on each line of a given file (the Suprime-CAM CCDs are noted to have very little bias pattern, so only the subtraction of the overscan median is required) and the overscan regions were then trimmed from the images. The flat field was then created using a total of 25 dithered object frames from across the 5 observed fields, with objects and regions vignetted by the auto-guider (AG) probe masked. With this flat field applied to the individual object frames, the astrometric distortion correction was performed (correcting for the telescope optics and the differential atmospheric dispersion) using a 5th order polynomial transformation \citep{Miyazaki2002}. The point spread function (PSF) was then matched across the images by applying a gaussian smoothing kernel to individual images. The sky background was then determined by interpolating over a mesh pattern. The image was divided into a grid of 64 x 64 pixel squares (corresponding to 12$''$.9 x 12$''$.9). A bilinear interpolation was used to determine the global sky background from this grid and the result was subtracted from the individual science frames. 

An initial astrometry solution was calculated using {\sc sdfred}, whereby the images were matched internally (i.e. using one of the science frames as a reference) and matched to a reference stellar catalogue. The images from each field were then co-added into stacked final images.

We applied the cosmic ray rejection using the rejected-mean algorithm, {\sc crreject}, from {\sc IRAF}.

\subsection{Object Detection and Photometry}

Extraction of sources from the images was performed using {\sc SExtractor} \citep{sextractor}. {\sc SExtractor} was run in dual image mode using the narrow band image as the detection image for each of the broad band images and the narrow band image itself. Photometric zero-points were calculated using the narrow-band observations of LTT~9491. As there is no NB497 narrow band standard photometry available for the standard star, we assumed the narrow band magnitude to be equal to the quoted $V$ band magnitude of $V=14.06$ (the $B-V$ colour of this star is $+0.03$). The zero-point magnitude for our observations given a 1~s exposure was $M_{\rm ZP}=24.34$, which equates to $M_{\rm ZP}=32.04$ for our 1200~s exposures. Taking into account atmospheric absorption for each field, the airmass ranged between $\approx1.1-1.5$. Taking an extinction coefficient of $k=0.12$ gives a range of extinction corrections of $\approx0.13-0.19$. The resulting individual zero-points are given in Tab.~\ref{tab:NB-Imaging}.

Aperture magnitudes were measured using an aperture diameter of $3.0''$ (approximately twice the seeing FWHM of the broad band data). Total magnitudes were measured using MAG\_AUTO.  $1\sigma$ magnitude depths were estimated using the errors calculated in $3.0''$ diameter apertures in {\sc SExtractor}. 

These initial catalogues were used for the selection of targets for the VLT VIMOS observations described in the following section.

\subsection{Improved Astrometry, Object Detection and Photometry}
\label{sec:improved_astrometry}
The initial astrometric solution was discovered to be inaccurate during the analysis of the VIMOS spectroscopic data (as described in Sec.~\ref{sec:vimos}). We therefore re-calibrated the astrometry and produced improved photometric catalogues for all of our fields.

This re-calibration of the astrometric solution was performed using {\sc scamp} \citep{scamp} with each stacked narrow band image in conjunction with the associated VLRS broad band images, thus providing a consistent astrometric solution across the narrow and broad band images. Since we applied the geometric transformation to all images, the object positions in each broad-band image  were matched with objects in the NB band. The distortion correction in the reduction process corrected the geometric distortion and we  obtained  good astrometry to better than $\pm0.2''$ rms over the image. Astrometry was made based on USNO at ESO catalogue and $\sim$ 1000 stars identified in the stacked images. The position of USNO  objects were approximately uniformly distributed over the entire stacked images. The absolute coordinates of our objects were obtained from these USNO objects. The images were then transformed based on the {\sc scamp} solution using {\sc swarp} \citep{swarp}.

We then produced two sets of catalogues based on these distortion-corrected images, both sets using SExtractor in dual image with one set using the narrow-band images as the detection image and the second set using a $\chi^2$ combined image as the detection image. The $\chi^2$ combined image was produced using {\sc swarp} and in each field combines the narrow band and $B$ and $R$ (or $V$) images in an optimal way for source detection \citep[see][for details]{swarp}. 

\section{Spectroscopic Observations}
\label{sec:vimos}

\subsection{Overview}

We have made spectroscopic follow-up observations of the LAE candidates with the VIsible Multi-Object Spectrograph (VIMOS)  \citep{2003SPIE.4841.1670L} on the Very Large Telescope (VLT). The VIMOS focal plane is divided into 4 quadrants, each measuring $7'\times8'$ (with gaps of $\sim2'$ between adjacent detectors). We observed with the HR\_Blue grism, allowing us to observe $\sim 20-30$ objects per detector with a spectral coverage of $4150~\AA<\lambda<6000~\AA$ and a resolution of $R\sim2,050$ (given our $1''$ slit-width).

VIMOS Mask Preparation was made by using VIMOS Mask Preparation Software (VMMPS)\footnote{http://www.eso.org/sci/observing/phase2/SMGuidelines/VMMPS.html} and the observations were conducted at the end of August to the beginning of September, 2011
(part of the observing run ESO-ID 086.A-0520 B, P.I. H. Francke).

The observations were taken in 3 of our 5 LAE fields: QSO B2359$+$068; QSO J0124$+$0044; and LBQS 0301$-$0035. Details of the observations are given in Table~\ref{tab:VIMOS-obs}.

\begin{table*}
\centering
\caption[Details of VIMOS LAEs observation]{Details of the VIMOS LAE observations.}
\begin{tabular}{lcccccc}
\hline
Field           & R.A.          & Dec.          & Constraints             & $N_{\rm LAE}$  & Seeing  & Integration time \\
                & \multicolumn{2}{c}{(J2000)}   &                         & (Sel/Obs)  & ($''$)  & (hours)          \\ 
\hline
QSO B2359$+$068 &  00:02:11.457	& +07:15:32.25  & $(R-m_{\rm NB497})>0.2$ & 304/37     & 1.0     & 2.0  \\
                &               &               & $(B-m_{\rm NB497})>0.9$ &            &         &      \\
QSO J0124$+$0044&  01:24:36.236	& +00:51:07.19  & $(V-m_{\rm NB497})>1.1$ & 141/45     & 1.0     & 4.3  \\
                &               &               & $(B-m_{\rm NB497})>1.4$ &            &         &      \\
LBQS 0301$-$0035&  03:03:10.208	& -00:16:20.98  & $(R-m_{\rm NB497})>0.8$ & 214/46     & 1.0     & 3.3  \\
                &               &               & $(B-m_{\rm NB497})>1.4$ &            &         &      \\
\hline
\end{tabular}
\label{tab:VIMOS-obs}
\end{table*}	

\subsection{Initial photometric target selection}
\label{sec:initsel}

\begin{figure}	
\centering
\includegraphics[width=\columnwidth]{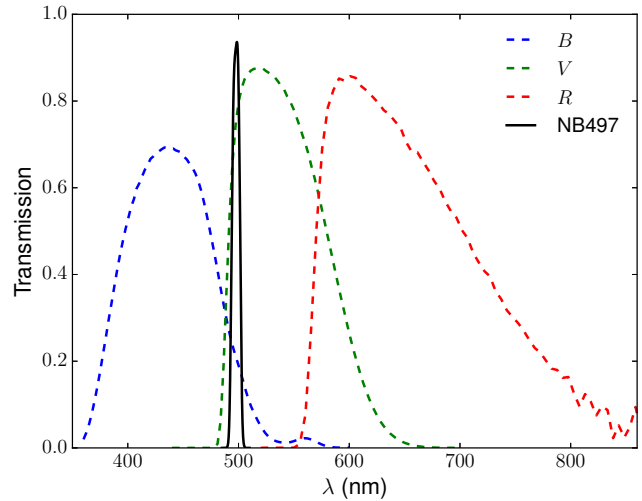}
\caption[The transmission curve of the filters]{The transmission curve of the filters. The solid line indicates the narrow-band filter, NB497. The dashed lines indicate broad-band filters, $B$, $V$, and $R$ from left to right respectively.}
\label{fig:trans_curve}
\end{figure}

LAEs were selected for the VLT VIMOS  spectroscopic observations using the narrow band 497~nm and broad band $B$, $V$ and $R$ band data already described. The transmission curves for each of these are shown in Fig.~\ref{fig:trans_curve}. For this selection we used a narrow band magnitude cut of $m_{\rm NB497}\leq26$ in combination with $(B-m_{\rm NB497})$ and $(V-m_{\rm NB497})$ or $(R-m_{\rm NB497})$ colour cuts. These colour cuts were tailored individually to each field in order to optimise the slit allocations in the VIMOS fields targeted. The colour cuts used in each field are given in Table~\ref{tab:VIMOS-obs}, along with the total number of objects selected (across the whole $\approx0.5^\circ\times0.5^\circ$ field in each case) and the number of these that were targeted in the VIMOS masks (i.e. over an area of $\approx16'\times16'$).

Different groups use different selection criteria to search for LAE candidates photometrically. For example: narrow-band and combined broad-band observations (e.g. $\left(B+R\right)/2$); one narrow-band filter and one broad-band filter; and one narrow-band filter and multiple individual broad-band filters. We choose the latter approach, using a two-colour approach which applied $R-{\rm NB497}$ colour and $B-{\rm NB497}$ colour, similar to those in \citet{Fynbo2003, Nilsson2007, 2008ApJS..176..301O}.

Following \citet{2008ApJS..176..301O}, we started from a colour cut of $(V-m_{\rm NB497})=1.2$, which equates to a cut in constant rest-frame equivalent width of ${\rm EW}_0\approx45$~\AA\ (or ${\rm EW}_0\approx65$~\AA\ when taking the Ly$\alpha$ Gunn-Peterson trough into account - see \citealt{2008ApJS..176..301O}). For our fields containing $R$ band data, but no $V$ band data, we took into account the mean $(V-R)$ colour difference in the $22<m_{\rm NB497}<25$ galaxy population giving an equivalent $R$-band cut is $(R-m_{\rm NB497})=0.8$. In addition to this, we also applied a cut using the $B$ band of $B-{\rm NB497}>1.3$. In order to maximise the slit allocations in the VIMOS masks, we then allowed these colour constraints to be relaxed. The resulting colour cuts used for the VIMOS target selection in each field are given in Tab.~\ref{tab:VIMOS-obs}. These colour cuts were combined with a magnitude limit of $m_{\rm NB497}\leq26$.

\subsection{Data reduction}

The reduction of the spectroscopic data was performed with the VIMOS pipeline \textsc{esorex} packages\footnote{http://www.eso.org/sci/software/cpl/esorex.html}. The main procedure includes creating master calibration data, reducing  science frames, and extracting objects. Following the pipeline manual, we firstly created a master-bias with the recipe {\it vmbias}. An output master-bias, was then used in  the reduction of the flat field, arc lamp, and scientific exposures. The next step was using the recipe {\it vmmoscalib} to  process  flat field exposures and arc lamp exposures. We used the {\it vmmosscience} recipe to process science frames with the cosmic ray rejection applied  at this point in the process. We then combined the images for each field using the {\sc iraf} {\it imcombine} task. 

The object extraction is made by applying an optimal extraction algorithm \citep{Horne1986}. The wavelength calibration was performed using the input wavelength calibration and sky lines, and sky background subtracted.  We found diagonal stray light interfered with the wavelength  calibration and sky subtraction in quadrants 3 and 4. To fix this, we isolated the brightest part of this light from the flat lamp, fitted the smooth pattern of the flat and subtracted it out.

\subsection{LAE identification}
\label{sec:laeident}

\begin{table*}
\centering
\caption{Galaxies with confirmed Ly$\alpha$ emission from the VLT VIMOS observations. The top, middle, and bottom entries correspond to the galaxies in the fields of the quasars B2359$+$0653, J01214$+$0044, and Q0301$-$0035 respectively}
\label{tab:vimoslaes}
\begin{tabular}{lcccccc}
\hline
ID  &  RA  &  DEC  &  z  &  $m_{497}$  &  $B-m_{497}$  &  $R-m_{497}$  \\ 
    & \multicolumn{2}{c}{(J2000)} &     & \multicolumn{3}{c}{(AB)} \\
\hline
VLRS J000223.50+071321.0   & 0.597932 & 7.222500 & 3.0939 & 23.48 & 2.44   & 1.50   \\ 
VLRS J000243.85+071021.5   & 0.682742 & 7.172648 & 3.1079 & 24.49 & 2.75   & 4.40   \\ 
VLRS J000222.05+071010.6   & 0.591878 & 7.169614 & 3.0543 & 24.81 & 1.96   & 2.91   \\ 
VLRS J000221.03+070915.6   & 0.587649 & 7.154351 & 3.1187 & 24.69 & 2.11   & 3.34   \\ 
VLRS J000150.11+072144.3   & 0.458817 & 7.362311 & 3.1112 & 25.57 & ---    & ---    \\ 
VLRS J000200.93+071939.3   & 0.503911 & 7.327593 & 3.0994 & 25.46 & ---    & 1.43   \\ 
VLRS J000143.31+071209.4   & 0.430497 & 7.202622 & 3.1015 & 25.55 & ---    & 1.40   \\ 
VLRS J000154.47+070912.2   & 0.476987 & 7.153391 & 3.1052 & 24.17 & 3.99   & 2.28   \\ 

\hline
ID  &  RA  &  DEC  &  z  &  $m_{497}$  &  $B-m_{497}$  &  $V-m_{497}$  \\ 
    & \multicolumn{2}{c}{(J2000)} &     & \multicolumn{3}{c}{(AB)} \\
\hline
VLRS J012457.44+004648.0   & 21.239342 & 0.780015 & 3.1030 & 24.70 & 1.68   & 3.29   \\ 
VLRS J012451.71+004502.5   & 21.215461 & 0.750693 & 3.0520 & 25.18 & 2.15   & 1.30   \\ 
VLRS J012450.74+004352.3   & 21.211416 & 0.731202 & 3.0811 & 24.55 & ---    & 0.24   \\ 
VLRS J012445.90+004317.0   & 21.191265 & 0.721403 & 3.0706 & 25.23 & 1.81   & 1.12   \\ 
VLRS J012509.24+005710.7   & 21.288526 & 0.952986 & 3.1228 & 24.84 & 3.05   & ---    \\ 
VLRS J012503.13+005634.7   & 21.263065 & 0.942975 & 3.1104 & 23.71 & 3.22   & 3.21   \\ 
VLRS J012408.25+005656.9   & 21.034377 & 0.949161 & 3.0748 & 24.15 & 3.02   & 4.67   \\ 
VLRS J012416.71+005622.2   & 21.069636 & 0.939499 & 3.0672 & 25.20 & 3.21   & 4.52   \\ 
VLRS J012413.08+005529.5   & 21.054527 & 0.924861 & 3.0734 & 25.54 & 0.44   & 0.86   \\ 
VLRS J012408.69+005341.3   & 21.036244 & 0.894817 & 3.0930 & 24.94 & 2.60   & 3.74   \\ 
VLRS J012428.28+005249.8   & 21.117851 & 0.880499 & 3.0761 & 24.00 & ---    & 1.53   \\ 
VLRS J012406.88+004750.3   & 21.028689 & 0.797311 & 3.0792 & 24.39 & 2.04   & 2.39   \\ 
VLRS J012417.96+004733.7   & 21.074841 & 0.792696 & 3.0786 & 24.25 & 3.22   & 3.07   \\ 

\hline
ID  &  RA  &  DEC  &  z  &  $m_{497}$  &  $B-m_{497}$  &  $R-m_{497}$  \\ 
    & \multicolumn{2}{c}{(J2000)} &     & \multicolumn{3}{c}{(AB)} \\
\hline
VLRS J030326.84-001805.8   & 45.861860 & -0.301630 & 3.0951 & 25.96 & ---    & 0.14   \\ 
VLRS J030337.40-001959.3   & 45.905844 & -0.333158 & 3.1322 & 25.67 & 0.54   & 0.39   \\ 
VLRS J030319.11-002139.7   & 45.829661 & -0.361044 & 3.0646 & 25.89 & ---    & ---    \\ 
VLRS J030325.62-002350.5   & 45.856789 & -0.397361 & 3.0757 & 25.28 & 2.96   & ---    \\ 
VLRS J030326.31-002400.2   & 45.859633 & -0.400077 & 3.0541 & 24.82 & 1.48   & 1.05   \\ 
VLRS J030333.31-001047.6   & 45.888794 & -0.179914 & 3.0746 & 24.57 & 2.07   & ---    \\ 
VLRS J030256.04-001007.2   & 45.733499 & -0.168692 & 3.0802 & 24.32 & 2.47   & 2.51   \\ 
VLRS J030248.36-001022.7   & 45.701519 & -0.172993 & 3.0943 & 25.52 & ---    & -0.01  \\ 
VLRS J030241.28-001140.6   & 45.672003 & -0.194622 & 3.1031 & 25.28 & 3.09   & ---    \\ 
VLRS J030244.58-001341.8   & 45.685771 & -0.228293 & 3.0730 & 23.87 & 1.52   & 1.21   \\ 
VLRS J030254.08-001414.6   & 45.725362 & -0.237397 & 3.0854 & 24.08 & 3.67   & 2.24   \\ 
VLRS J030244.33-001911.9   & 45.684716 & -0.319997 & 3.1251 & 24.37 & 1.41   & 0.62   \\ 
VLRS J030257.61-002058.5   & 45.740049 & -0.349601 & 3.0939 & 24.76 & ---    & 3.26   \\ 
VLRS J030257.31-002301.0   & 45.738794 & -0.383629 & 3.1019 & 25.62 & 1.90   & 0.82   \\ 

\hline
\end{tabular}
\end{table*}

We found significant emission within the wavelength range of the narrow band filter in 8 of the 23 candidates in the QSO B2359$+$0653 field, 13 of the 45 in the QSO J0124$+$0044 field, and 14 of the 46 in the LBQS 0301$-$0035 field. These are listed in full in Tab.~\ref{tab:vimoslaes}, whilst the reduced 1D spectra (with 2D spectra inset) are shown in Fig.~\ref{fig:Q2359-vimos-spectra}, Fig.~\ref{fig:J0124-vimos-spectra} and Fig.~\ref{fig:Q0302-vimos-spectra} for the three separate fields. In each figure, the solid black line shows the reduced 1D spectrum, the shaded grey region shows the normalised NB497 filter response and the inset image shows the 2D spectrum within $\pm60$~\AA\ of the peak of the NB497 filter response curve.

Given all our detections are single line detections, the possibility exists that some of these could be [OII] emission from faint $z\approx0.33$ galaxies. The spectral resolution of our observations is marginally sufficient to discern the double peaked emission inherent in the OII double lines (the resolution is $\sim$ 2.5 \AA\, while the [OII] doublet is about 2.7 \AA). Although the resolution is very close to the peak separation, even in the event of the doublet being smoothed out in the spectrum by the instrument response, the presence of the [OII] doublet would lead to a significantly broad line. We have analysed each emission line, plotting the doublet separation over each and evaluating the likelihood of each line being [OII] emission. In most cases ($\gtrsim90\%$), clearly shows no sign of being a doublet (i.e. there is no double peak and the single peak emission is not broad enough to be the two [OII] lines convolved into one via the instrument response). For the remaining $\lesssim5\%$, we cannot discount the possibility of the emission being low redshift [OII], but at the same time, none are clear-cut cases of [OII] emission. These ambiguous $\lesssim5\%$ are broad emission that show no significant detections of double peaks within the spectral noise. Indeed higher signal-to-noise and resolution observations would be needed to fully discern any double peaked nature to these broader lines. In addition, we note that even if double-peaked emission were detected, the complex nature of the escape of Ly$\alpha$ photons from galaxies can also produce double emission peaks. In all we find no strong evidence of any double emission in any of the detected emission lines and conclude that $\gtrsim95\%$ of these lines are Ly$\alpha$ emission at $z\approx3.06$.

The success rate of the observations were somewhat low in part due to the relaxed constraints used to select candidates. Additionally, the slit mask for the second quadrant in the QSO B2359$+$0653 field was not properly aligned in its mount leading to no detections in that quadrant. However, in addition to this, a poor astrometric solution was found to have contributed to the low success rate. The astrometric errors for the images created two issues: 1. in the QSO B2359$+$0653 field, a number of targets were not correctly aligned within the slits during the observations; and 2. inconsistencies between the astrometry between the narrow band and broad bands led to incorrect broad band photometry measurements. Apertures placed to measure the photometry in the broad band images were thus offset from the intended targets in some cases, leading to incorrectly faint magnitudes being measured and objects with no emission being promoted into the selection criteria.

Following the spectroscopic observations, we thus recalculated matched astrometric solutions for all the data as described in Sec.~\ref{sec:improved_astrometry}.

\subsection{Selection efficiency and optimised selection criteria}
\label{sec:selec-eff}

\begin{figure*}	
   	\centering
   	\includegraphics[width=\textwidth]{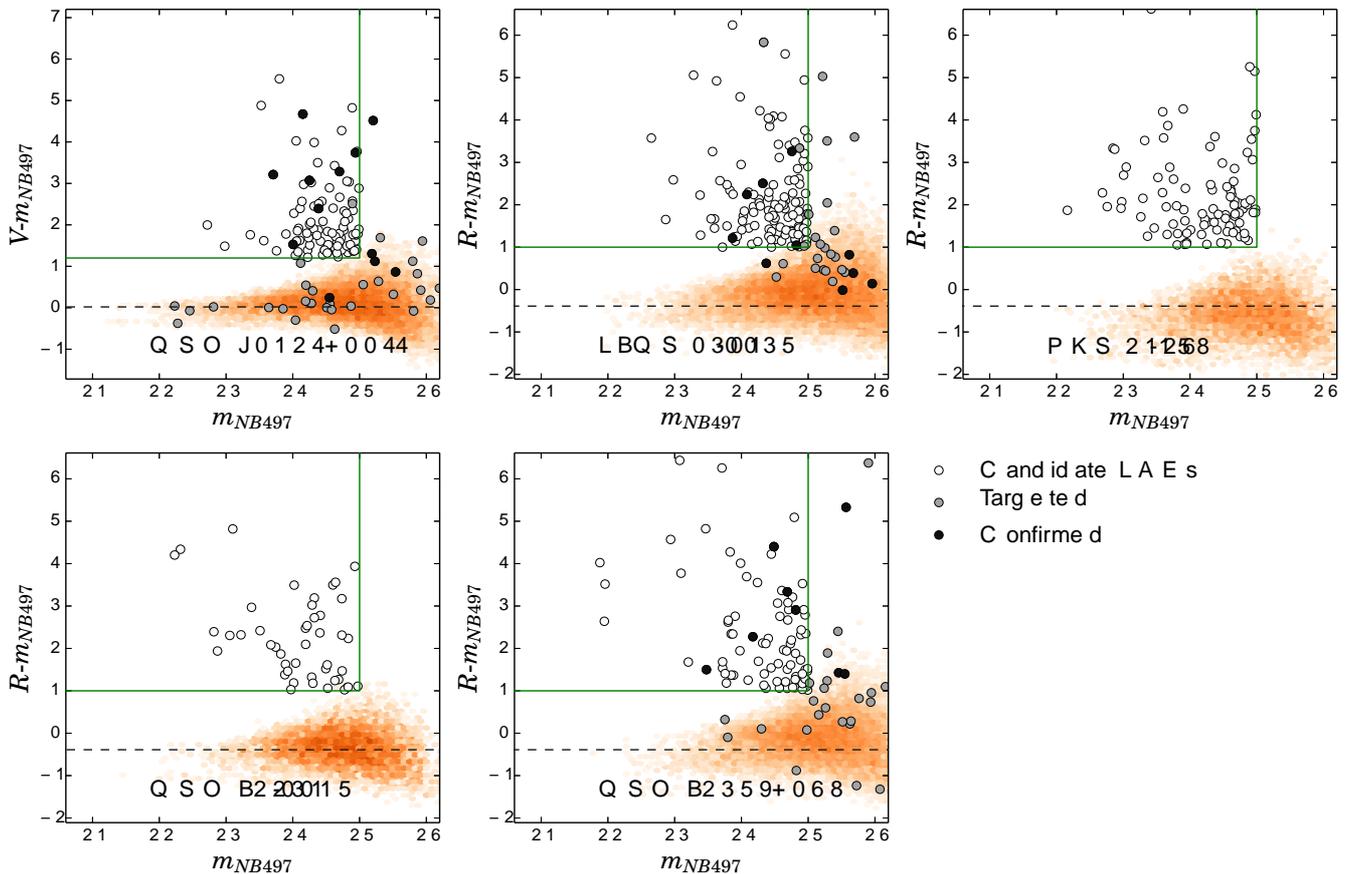}
	\caption{Colour magnitude diagrams showing $V-m_{\rm NB497}$ versus $m_{\rm NB497}$ for the QSO J0124$+$0044 field and $R-m_{\rm NB497}$ versus $m_{\rm NB497}$ for the LBQS 0301$-$0035, PKS 2126$-$158, QSO B2231$-$0015, and QSO B2359$+$068 fields. The shaded 2D histogram shows the general galaxy population. Grey and black filled circles show LAE candidates observed with VLT VIMOS, the grey showing those with no spectroscopic detection and the black showing those confirmed as $z=3.06$ LAEs. The region to the upper left of the solid horizontal and vertical bounding lines give the updated selection region, whilst the open circles show LAE candidates selected based on these bounds. Note that the objects identified as being targeted with VIMOS that lie within the main galaxy population are those that were subject to significant photometric errors due to the astrometric mis-alignment of images (see Sec.~\ref{sec:laeident}).}
   	\label{fig:colour-mag-R}
	\end{figure*}

\begin{figure*}	
   	\centering
   	\includegraphics[width=\textwidth]{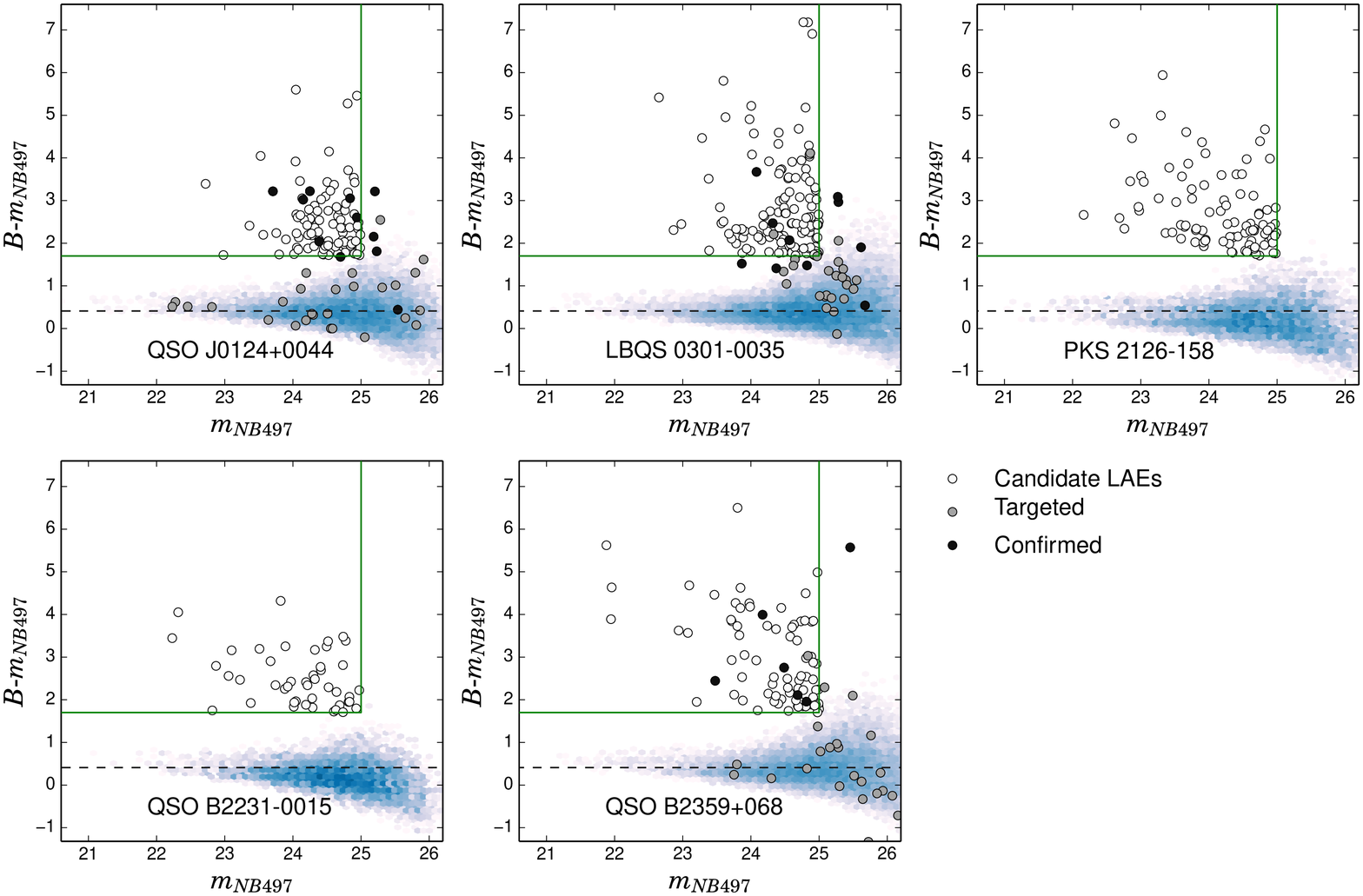}
	\caption{Colour magnitude diagrams showing $B-m_{\rm NB497}$ versus $m_{\rm NB497}$ for our five QSO fields: QSO J0124$+$0044, LBQS 0301$-$0035, PKS 2126$-$158, QSO B2231$-$0015, and QSO B2359$+$068. The shaded 2D histogram shows the general galaxy population. Grey and black filled circles show LAE candidates observed with VLT VIMOS, the grey showing those with no spectroscopic detection and the black showing those confirmed as $z=3.06$ LAEs. The region to the upper left of the solid horizontal and vertical bounding lines give the updated selection region, whilst the open circles show LAE candidates selected based on these bounds. Note that the objects identified as being targeted with VIMOS that lie within the main galaxy population are those that were subject to significant photometric errors due to the astrometric mis-alignment of images (see Sec.~\ref{sec:laeident}).}
   	\label{fig:colour-mag-B}
	\end{figure*}

The selection success rates are relatively low in our spectroscopic sample of LAE candidates (31\%). This is the result of the original criteria being comparatively flexible in order to allow the maximal number of slits placed per VIMOS field, whilst as discussed the astrometric calibrations were sub-optimal prior to the spectroscopic follow-up, causing errors in the photometry. In this section we therefore provide an optimised and uniform set of selection constraints, maximising the numbers of successfully identified LAEs based on our updated photometric catalogues and the spectroscopic observations.

First of all, we now use a narrow band magnitude cut of $m_{\rm NB497}\leq25.0$ (as opposed to $m_{\rm NB497}\leq26$ previously), as in all three fields there is significant scatter from the galaxy locus at fainter magnitudes than this due to the measurement uncertainties on the photometry. The converse of this is that the measured colours of many LAEs beyond this limit are scattered into the galaxy locus in the colour-magnitude diagram.

Secondly, we optimise the $V-m_{\rm NB497}$, $R-m_{\rm NB497}$ and $B-m_{\rm NB497}$ colour cuts at $m_{\rm NB497}\leq25.0$ in order to maximise the numbers of LAEs found and minimise the number of interlopers given the spectroscopic observations. We also apply the constraint that the three colour criteria correspond to the same equivalent width constraint. For this we simulate LAE spectra by assuming a $\beta=-1.4$ power-law slope \citep[e.g.][]{2012ApJ...749....4B} combined with a range of equivalent Ly$\alpha$ emission lines. We find that taking colour constraints of $V-m_{\rm NB497}\geq1.2$, $R-m_{\rm NB497}\geq1.0$ and $B-m_{\rm NB497}\geq1.4$ selects 18 out of the 20 spectroscopically confirmed $m_{\rm NB497}$ LAEs at $m_{\rm NB497}\leq25.0$, whilst missing 2 of these. The selection also includes 5 non-detections from the spectroscopic observations that may be contamination or weakly emitting $z\approx3.1$ LAEs (likely scattered into the selection region due to noise on the photometry). These colour cuts are equivalent to a rest frame equivalent width cut of $\approx45$~\AA, when assuming the simple $\beta=-1.4$ continuum power law slope. We do not however include the effect of the Gunn-Peterson trough and so, as discussed and modelled in \citet{2008ApJS..176..301O}, our nominal $\approx45$~\AA\ selection corresponds to a $\approx65$\AA\ cut when taking this into account.

Additionally, we also reject objects with {\sc SExtractor} flags greater than zero; mask low signal-to-noise regions of the images (primarily the image edges) to avoid noise contamination in these regions; and (in order to minimise the contribution of artefacts in the images) apply elongation ($e_{\rm SEx}$) and FWHM limits. By inspection of the images, objects with a measured elongation of $e_{SEx}>2.8$ are exclusively artefacts (primarily cosmic ray hits not removed by {\sc crreject}). In addition, objects with measured FWHM less than the measured image FWHM are also rejected (on inspection these are also exclusively artefacts and primarily remnant cosmic rays).

\begin{table}
\centering
\caption{Number of LAE Candidates from our selection in each field.}
\begin{tabular}{lcc}
\hline
Field           & $N_{\rm LAE}$ & $\rho_{\rm LAE}$ \\ 
                &               & (arcmin$^{-2}$) \\ 
\hline
QSO B2359$+$068   &  124          &  0.160           \\
QSO J0124$+$0044  &  139          &  0.186           \\
LBQS 0301$-$0035  &  170          &  0.201           \\
PKS 2126$-$158    &  140          &  0.158           \\
QSO B2231$-$0015  &  70           &  0.119           \\
\hline
All fields        &  643          &  0.167           \\
\hline
\end{tabular}
\label{tab:LAEcan}
\end{table}

The selection criteria are summarised as follows: 

\begin{itemize}
\item $20 < m_{\rm NB497} < 25$;
\item $(V - m_{\rm NB497}) > 1.2$ or $(R - m_{\rm NB497}) > 1.0$;
\item $(B - m_{\rm NB497}) > 1.4$;
\item ${\rm FLAG}_{\rm SEx}=0$;
\item $e_{\rm SEx}<2.7$;
\item ${\rm FWHM}\geq{\rm FWHM}_{\rm image}$.
\end{itemize}

These colour cuts are shown by the solid lines in Figs.~\ref{fig:colour-mag-R} and \ref{fig:colour-mag-B}. In each figure the galaxy population is shown by the 2d histogram, whilst selected LAE candidates are shown by open circles. The dashed horizontal lines show the position of the locus in $(V - m_{\rm NB497})$ or $(R - m_{\rm NB497})$ colour (Fig.~\ref{fig:colour-mag-R}) and $(B - m_{\rm NB497})$ colour (Fig.~\ref{fig:colour-mag-B}). The grey and black filled circles show LAE candidates observed with VLT VIMOS and are described further in Sec.~\ref{sec:vimos}. The seeing for each field (i.e. ${\rm FWHM}_{\rm image}$) is given in Tab.~\ref{tab:NB-Imaging}. As discussed, we have produced catalogues based on object detection in the narrow-band images alone and also in a $\chi^2$ combined image of the narrow and broad band images. We have run our selection on both of these sets of catalogues. In the results that we report here, we use the $\chi^2$-image detected catalogues. We have run all our analysis on the narrow-band image detected catalogues and find consistent results within the estimated uncertainties.

The numbers and sky densities of candidates in each field given by these criteria are given in Tab.~\ref{tab:LAEcan}. Taking only those spectroscopically observed targets that fall within the updated selection criteria, we find a success rate of $78\pm18\%$. The proportion of targets selected in this new sample that were also present in the original target selection is $\approx40\%$, thus calculating the success rate for our updated selection using the spectroscopic sample is arguably only representative of 40\% of the sample. However, the remaining 60\% of targets selected in the updated selection were missed in the original selection due to astrometric errors. The loss of targets due to the sub-optimal astrometric solution has no connection with intrinsic source properties (rather it is a function of the on-sky coordinates of the sources) and so should not introduce any biases between the 40\% that were selected and the 60\% that were not, so it would be reasonable to assume that the result for the 40\% that were included, holds for the 60\% that were missed due to astrometric issues.

Comparing to the success rate in other's work, \citet{2008ApJS..176..301O} identified line emitters from 60$\%$ of their targets at this redshift, whilst \citet{Fynbo2001,Fynbo2003} reported the spectroscopic follow-up success rate of 75 - 90$\%$ for $z\sim3$ LAE surveys. 

\begin{figure*}
\centering
\includegraphics[width=\textwidth]{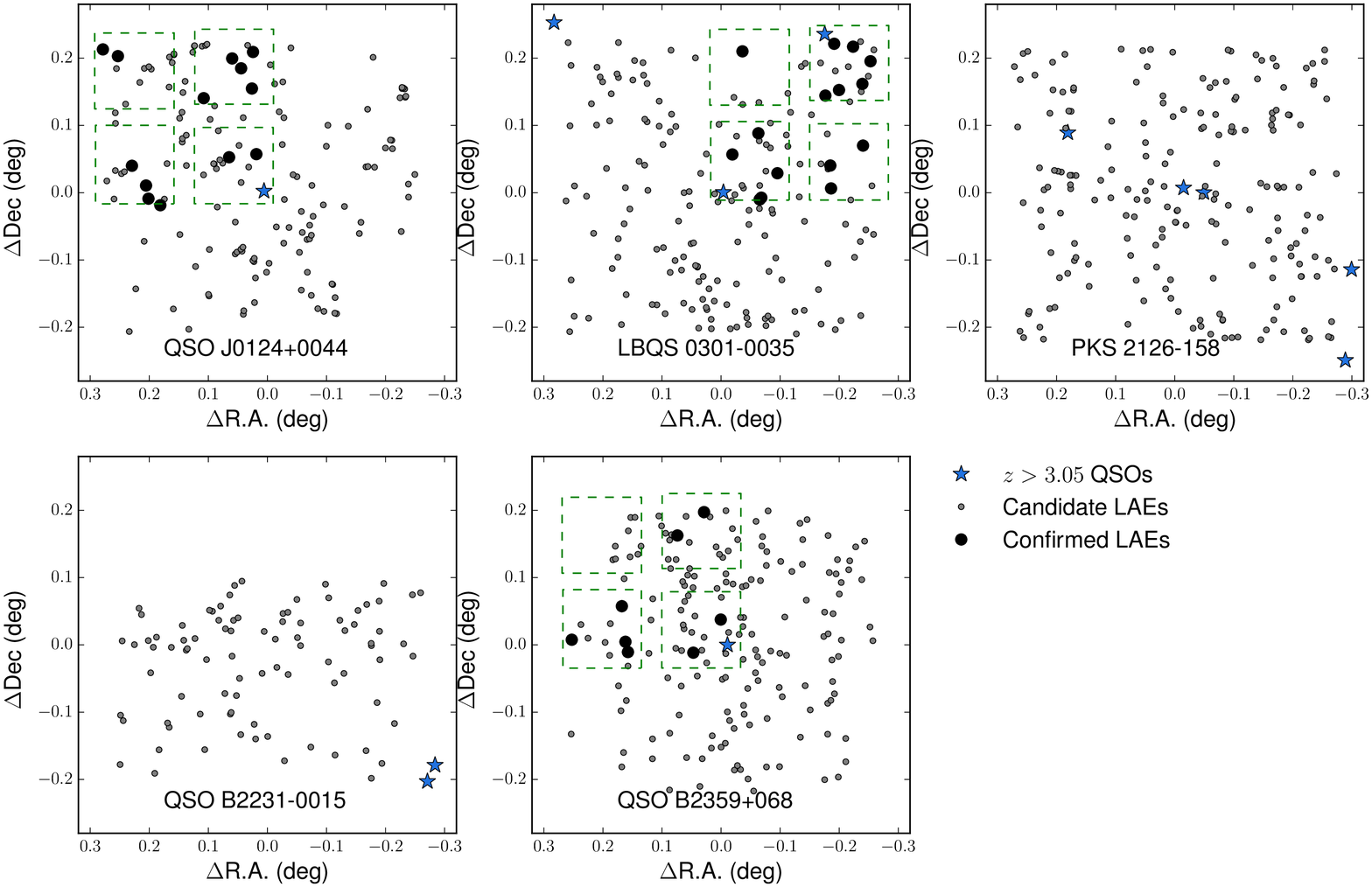}
\caption{The distribution of candidate (grey filled circles) and confirmed (black filled circles) LAEs in the five survey fields. Known QSOs at the LAE redshifts or higher (i.e. $z>3.05$) are also plotted for reference (stars). The dashed green boxes show the areal coverage of the VIMOS quadrants for the spectroscopic observations (note that the fields PKS~2126-158 and QSO~B2231-0015 fields have not been observed spectroscopically). The central QSO in the QSO~B2231-0015 field is not plotted as it has a redshift lower than that of the LAE selection ($z_{\rm QSO}=3.027$).}
\label{fig:LAE-dis1}
\end{figure*}

The distribution of the selected LAE candidates in each field is shown in Fig.~\ref{fig:LAE-dis1}, with the same symbols as in previous plots (i.e. open circles for LAE candidates and filled grey and black circles for candidates observed with VLT VIMOS). Filled stars in Fig.~\ref{fig:LAE-dis1} denote the positions of background QSOs in the fields.

\subsection{Redshift and spatial distribution of LAEs}
\label{sec:nz}

\begin{figure}	
\centering
\includegraphics[width=80.mm]{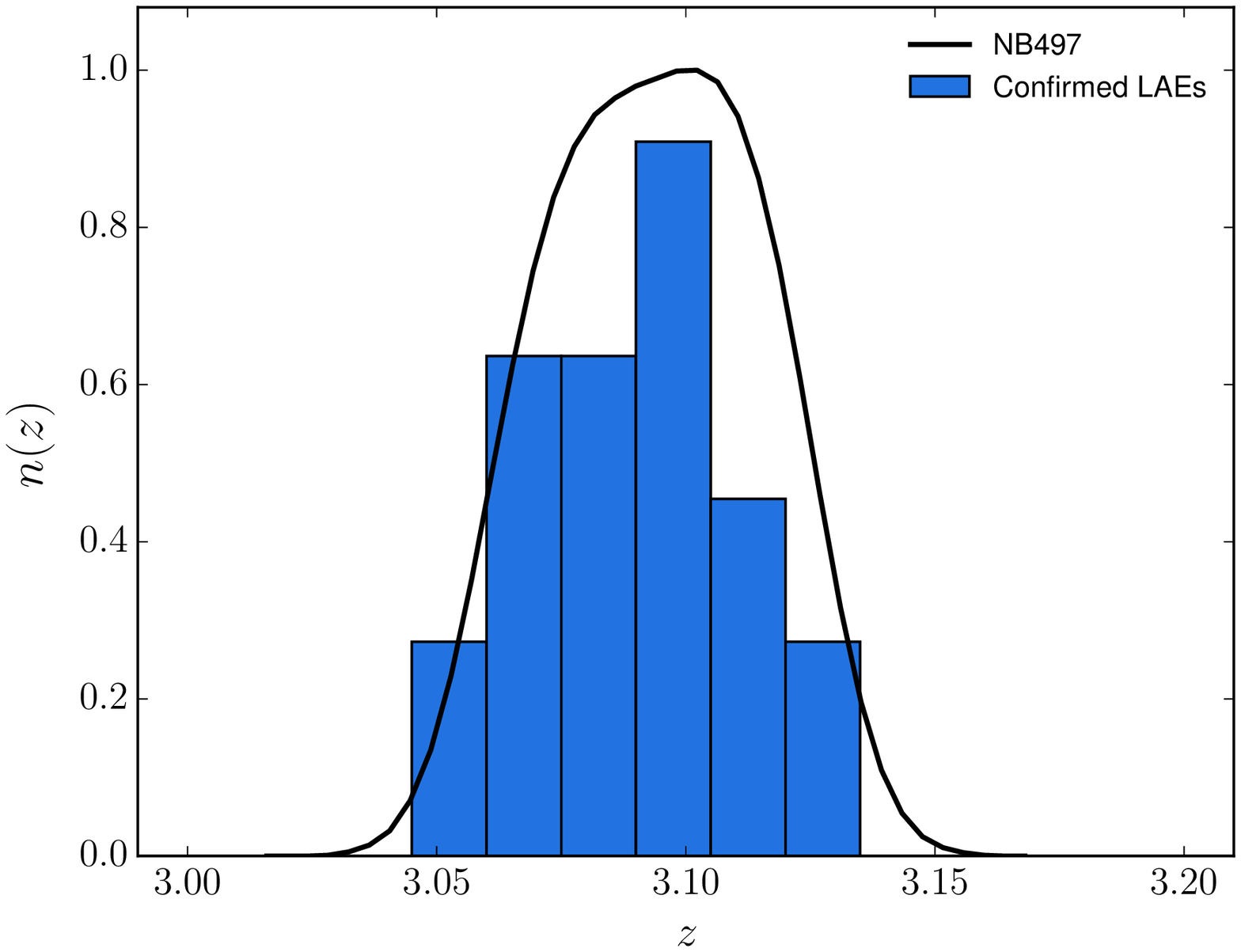}
\caption[Redshift distribution for LAEs]{Redshift distribution for LAE spectroscopic identification from VIMOS (normalised histogram) and the NB497 filter with wavelength converted to redshift (solid curve). The NB497 filter is normalised such that the peak is unity.}
\label{fig:LAE-nz}
\end{figure}

The redshift distribution, $n(z)$, for the VLT VIMOS confirmed LAEs is shown in Fig.~\ref{fig:LAE-nz} (solid blue histogram). For comparison, the throughput curve of the NB497 filter is also shown (solid black curve).

The survey sky coverage is illustrated in Fig.~\ref{fig:LAE-dis1}, where the distributions of candidate LAEs (small grey circles) and confirmed LAEs (large black circles) are shown for each field. Background $z\gtrsim3.1$ QSOs are also shown (filled-stars). These five fields cover a total area (after masking) of 1.07~deg$^2$.

Integrating along the normalised VIMOS redshift distribution and combining it with this total sky area gives an LAE survey volume of $2.51\times10^{5}~h^{-3}{\rm Mpc}^3$. Integrating instead the NB497 filter response curve gives a somewhat higher survey volume estimate of $3.35\times10^{5}~h^{-3}{\rm Mpc}^3$. In the analysis that follows, we take the volume given by the filter response curve given that the measured redshift distribution is somewhat under-sampled. For reference, a top-hat redshift distribution (between $z=3.05$ and $z=3.14$) gives a survey volume of $4.69\times10^{5}~h^{-3}{\rm Mpc}^3$.

\subsection{Selection Completeness}
\label{sec:completeness}

In order to estimate the selection completeness, we first estimate the percentage completeness as a function of source magnitude of the individual survey images. To do so, we simulate sources in the images in set magnitude bins. We attempt to detect the simulated sources using SExtractor, identically to the extraction process for the original data. For each magnitude bin, we simulate 100 sources, with the number recovered giving the fractional completeness in each bin. We avoid source confusion by only placing simulated sources in empty regions with no genuine sources found within $2.5\times$ the image FWHM. 

\begin{figure}
\centering
\includegraphics[width=\columnwidth]{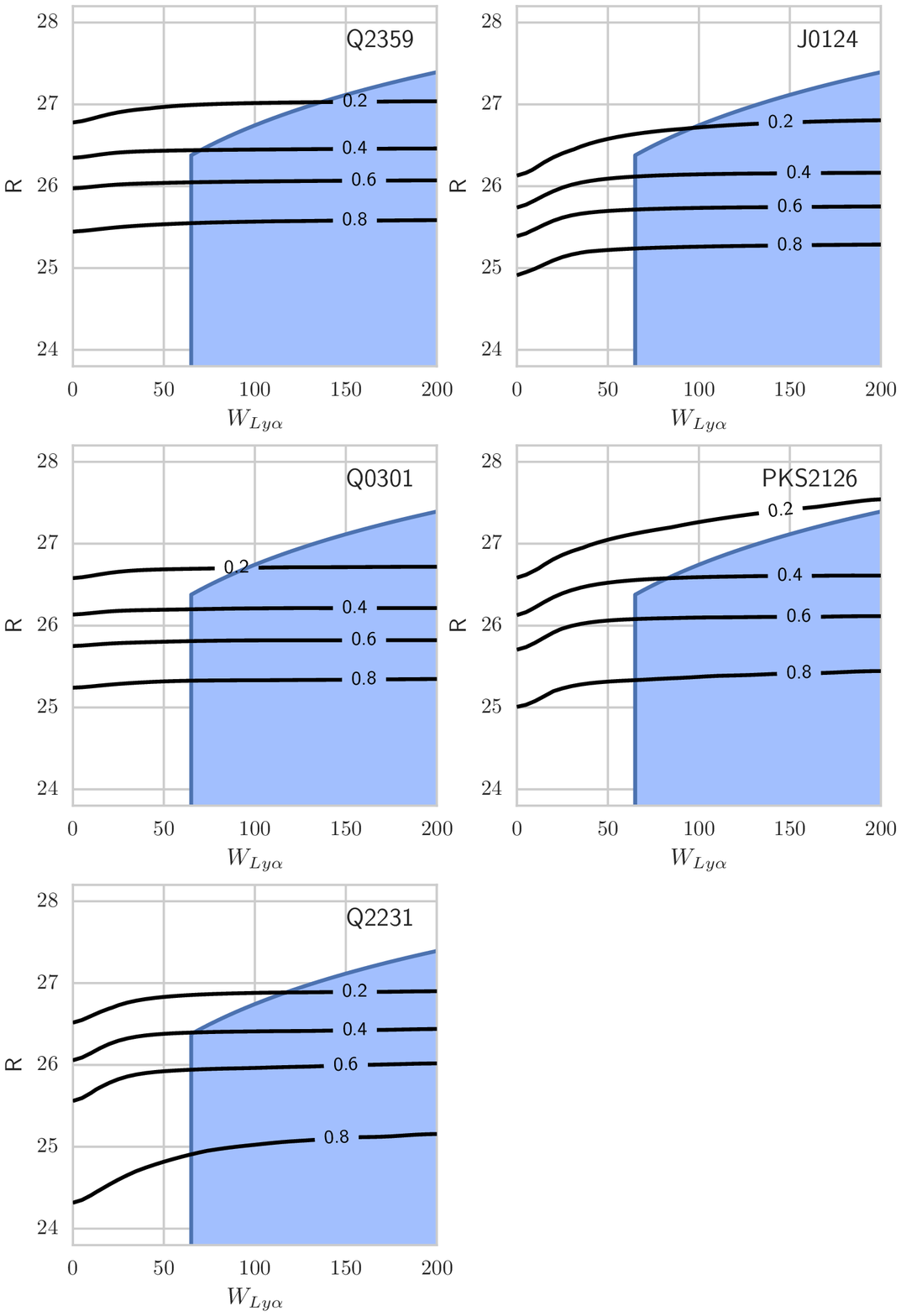}
\caption{Estimated selection completeness in the five observed fields based on simulated sources in the survey images. The shaded region shows the extent of our selection criteria in $R-$EW space, whilst the contour curves show the 20\%, 40\%, 60\%, and 80\% completeness levels.}
\label{fig:selcomp}
\end{figure}

We applied these detection rates with simulated LAE spectra with a range of equivalent widths and assuming a spectral slope of $\beta=-1.4$ \citep[e.g.][]{2012ApJ...749....4B}. The results of this simulation for each field are shown in Fig.~\ref{fig:selcomp}. Each panel shows one of our five individual fields, with the shaded regions showing the range of our LAE selection in $R$-EW space. The contour curves show the 20\%, 40\%, 60\%, and 80\% completeness levels.

Overall, we find good consistency across our five fields in terms of the completeness profile. At fainter continuum luminosities (i.e. $R\gtrsim25.5$), we find that we increasingly miss the high equivalent width range of our selection with all 5 fields falling to $\approx20\%$ completeness at $R\approx27$.

\section{Number counts \& Luminosity Function}
\label{sec:lumfunc}

\subsection{Narrow band number counts}
\label{sec:nb-lf}

\begin{figure}
\centering
\includegraphics[width=80.mm]{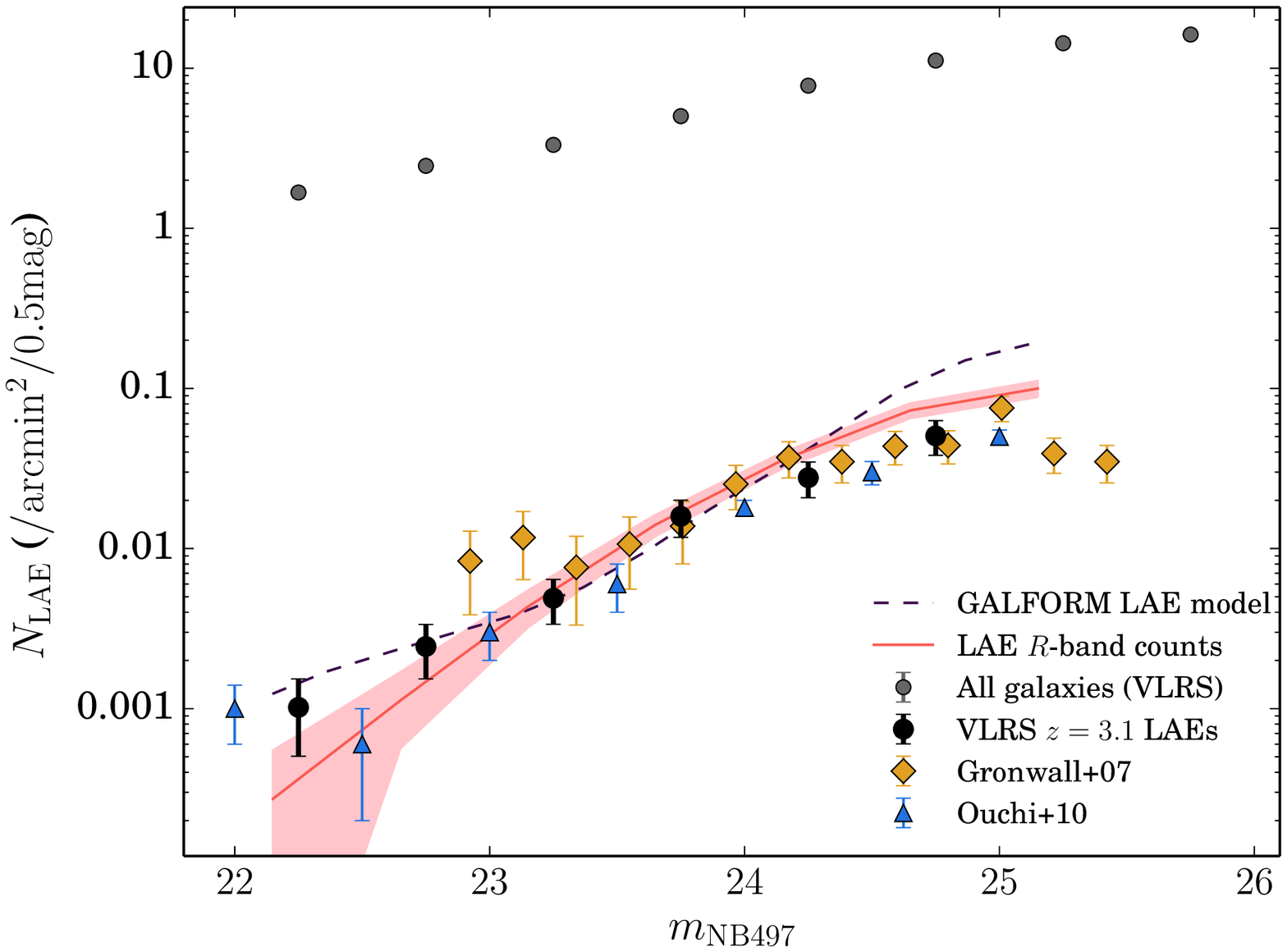}
\caption{The number counts of photometrically detected LAEs based on the selection criteria given in Sec.~\ref{sec:selec-eff} (filled black circles). The diamonds and triangles show number counts for comparable LAE selections by \citet{Gronwall2007} and \citet{2008ApJS..176..301O} respectively. Small filled grey circles show the number counts of all objects detected in the narrow band imaging. The solid curve shows model LAE number counts from the GALFORM galaxy formation model. The shaded curve shows the LAE $R$-band counts adapted from Fig.~\ref{fig:LF-LAEs-LBG} with a shift of $\Delta m=-1.6$ to overlay the narrow band counts.}
\label{fig:NB-density}
\end{figure}

Using the updated selection criteria given in the previous section, we calculate the galaxy number counts for the LAE sample as a function of narrow band magnitude. This is shown in Fig.~\ref{fig:NB-density}, where the filled black circles show our results. The error bars show field-to-field error estimates on the points. For reference, we also show the NB497 number counts for all sources in our fields (grey circles). 

For comparison, we show the $z\approx3.1$ narrow band number counts published by \citet[][diamonds]{Gronwall2007} and \citet[][triangles]{2008ApJS..176..301O}. \citet{Gronwall2007} conducted a 0.28~deg$^2$ survey at $z\approx3.1$ in the Extended Chandra Deep Field South (ECDFS). Their selection is based on a measured colours derived from the NB5000 filter and a broad band image constructed using a combination of images taken using the $B$ and $V$ band filters. Their criteria were thus:$m_{\rm NB5000} - m_{B+V} < 1.03$, with a narrow-band magnitude limit of $m_{\rm NB5000}\leq25.4$. This gave a sample of 162 galaxies with rest-frame equivalent widths of $>20$~\AA\ (approximately 80 \AA\ in the observer's frame). The \citet{2008ApJS..176..301O} selection criteria are based on a cuts of $V-m_{\rm NB503}>1.2$ and $B-V >0.5$. They detected 356 photometric LAEs, with rest-frame equivalent widths of $\geq65$~\AA, over a $\simeq$~1~deg$^2$ area. 

A direct comparison of the \citet{Gronwall2007} and \citep{2008ApJS..176..301O} number counts of $z\sim3.1$ LAEs is shown by \citet{Ciardullo2012}. \citet{Ciardullo2012} extended the study of \citet{Gronwall2007} by re-imaging the ECDF-S with a 57 \AA\ FWHM nearly top-hat filter centred at $\approx5010$~\AA. They found a total number of 360 $z\approx3.1$ LAEs, a subset of which are also selected by \citet{Gronwall2007}. They then reproduced the luminosity function  and concluded that their result is statistically identical to values in \citet{Gronwall2007} and \citet{2008ApJS..176..301O}, once differences in the filter FWHM and EW selection constraints were accounted for.

The VLRS LAE number counts are consistent with the number counts of \citet{Gronwall2007} at the $\approx1\sigma$ level, and marginally higher than the counts of \citet{2008ApJS..176..301O} (although still only at the $\approx1-2\sigma$ level). Our plotted number counts include a factor of 0.78 to account for the number of non-detections when our sample was observed spectroscopically. This is perhaps a conservative estimate for the success of our selection criteria in identifying LAEs, given that some fraction of those non-detections could have been LAEs below the detection threshold of the spectroscopic observations. In the most extreme case, i.e. if all the non-detections were in fact genuine LAEs, the VLRS LAE number counts shown in Fig.~\ref{fig:NB-density} could be increased by a factor of $\approx25\%$. This would perhaps cause some tension with the counts of \citet{2008ApJS..176..301O} at the $2-3\sigma$ level. 

For context in comparing the individual results, the equivalent width limits for each survey are: 65~\AA\ (this paper), 20~\AA\ \citep{Gronwall2007}, and 65~\AA\ \citep{2008ApJS..176..301O}. \citet{Ciardullo2012} suggest that the lower equivalent width limit of \citet{Gronwall2007} should measure $\approx10\%$ higher numbers of LAEs compared to \citet{2008ApJS..176..301O}, and hence our own sample. In addition to the different EW cuts, given the narrower filter used by \citet{Gronwall2007}, one would expect to measure lower counts by a factor of $\frac{2}{3}$ given the smaller volume coverage this entails. \citet{Ciardullo2012} do not explicitly state that the difference in filter width is included in their prediction of a difference of 10\% in the numbers of LAEs between the samples, however their estimate would seem somewhat low if it is not (for example when compared to the equivalent width distribution of LBGs measured by \citealt{shapley03}).

Although our main aim has been to check the representiveness of our LAE counts as a prelude to using the LAE for clustering studies, we now briefly compare our counts to a recent theoretical model. Shown in Fig.~\ref{fig:NB-density} therefore is the \citet{Orsi2012} model for LAEs updated to the latest variant of the GALFORM semi-analytical model of galaxy formation \citep{2015arXiv150908473L}. This model for LAEs makes use of a Monte Carlo Ly$\alpha$ radiative transfer code to compute the escape fraction of LAEs at different redshifts. The GALFORM model is calibrated to match a number of observational data, mostly at $z=0$ (e.g. optical and NIR LFs, $M_{\rm BH}-M_{\rm bulge}$ relation, morphological fractions), but also sub-mm, FIR counts and UV luminosity functions at high redshift. A radiation transfer model was then used to model the Ly$\alpha$ escape fraction, based on an outflow model which is itself dependent on the individual galaxy properties. This radiation transfer and outflow model (which in this case consists of an expanding thin shell of material) was tuned to match the luminosity function of \citet{2008ApJS..176..301O}. Given the agreement between our own results and those of \citet{2008ApJS..176..301O}, it is no surprise to find the model successfully reproduces our own narrow band number counts at $m_{NB497}\lesssim24.2$, although the agreement worsens at fainter magnitudes (as it does with the \citealt{2008ApJS..176..301O} results). Even to the extent that they agree, the tuning of the model to the luminosity function of \citet{2008ApJS..176..301O} means that the observed LAE number counts are therefore not a test of the model. This and the disagreement at faint fluxes must be borne in mind when comparing the LAE clustering data with the model in Sec.~\ref{sec:clustering_results}.

\subsection{Continuum luminosity function}

\begin{figure}	
\centering
\includegraphics[width=80.mm]{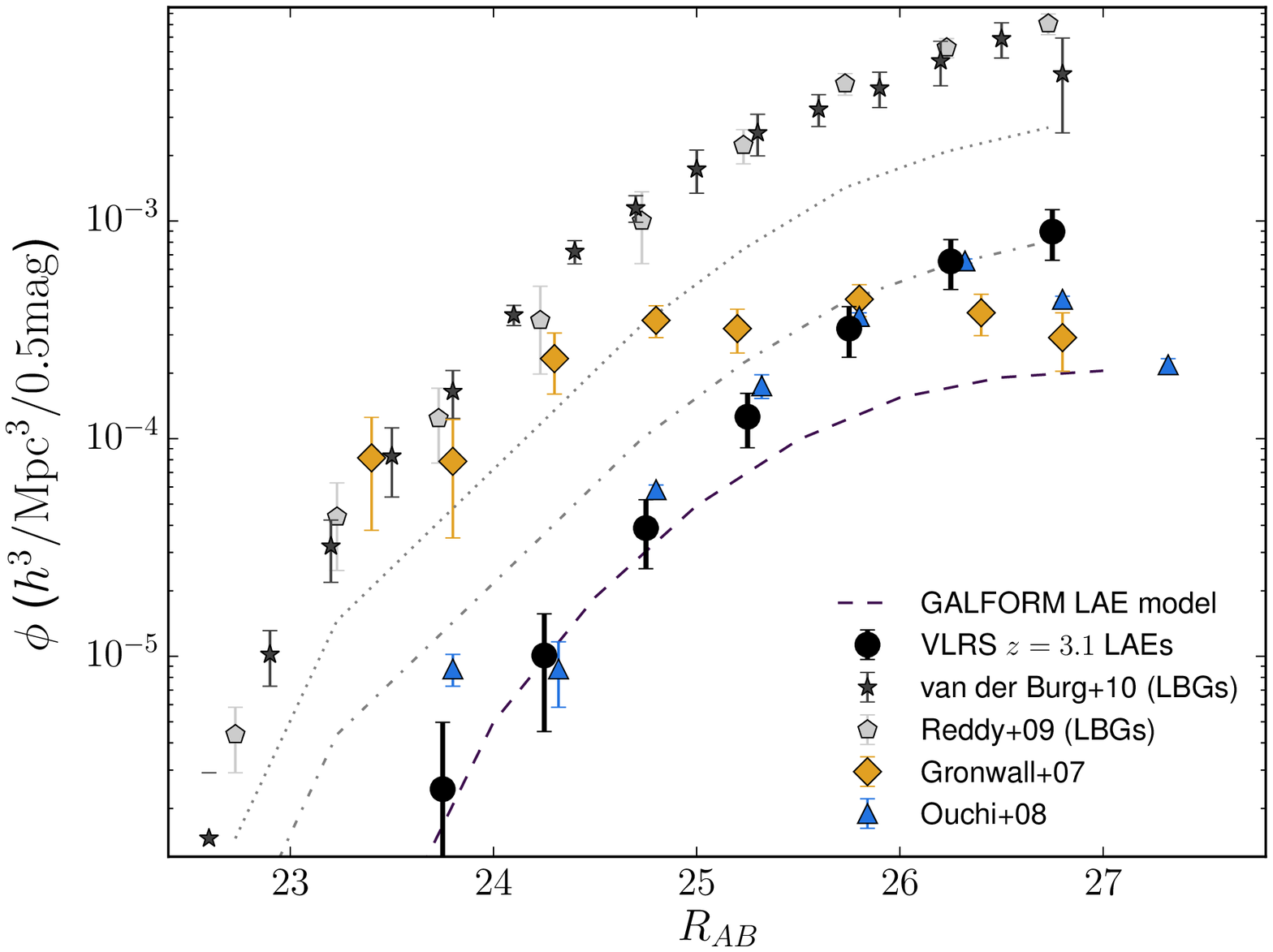}
\caption[Luminosity function of LAEs]{The $R$-band luminosity function of the VLRS $z=3.1$ LAEs (filled circles). For comparison, we also show the equivalent LAE luminosity functions from \citet[][diamonds]{Gronwall2007} and \citet[][triangles]{2008ApJS..176..301O}. Also shown is the observed $r$-band luminosity functions of $z\sim3$ LBGs from \citet[][pentagons]{2009ApJ...692..778R} and from the CFHT fields of \citet[][stars]{vanderBurg2010}. The dotted and dash-dot curves show the \citet{2009ApJ...692..778R} luminosity function reduced by a factor of 3 and a factor of 10 respectively, whilst the dashed curve shows the GALFORM model result.}
\label{fig:LF-LAEs-LBG}
\end{figure}

We estimated the $R$-band continuum luminosity function of our sample of LAEs. As in \citet{Gronwall2007}, we use the filter curve to define the survey volume used for the luminosity function calculation, which gives a volume of $3.35\times10^{5}~h^{-3}{\rm Mpc}^3$. The resulting continuum luminosity function evaluated across all 5 fields (and scaled in accordance with the estimated 78\% success rate of our sample) is shown in Fig.~\ref{fig:LF-LAEs-LBG} (filled black circles). Plotted points are corrected for the incompleteness as a function of $R$ magnitude estimated in Sec.~\ref{sec:completeness}. The error bars give the uncertainty based on field-to-field estimates using our five imaging fields in combination with the uncertainty introduced by the estimate of the same success rate.

We show our results in comparison to those of \citet[][diamonds]{Gronwall2007} and \citet[][triangles]{2008ApJS..176..301O}. Our results are consistent with those of \citet{2008ApJS..176..301O} within the calculated errors to a magnitude of $R\approx26$. In contrast, our luminosity function (and that of \citealt{2008ApJS..176..301O}) is only consistent with that of \citet{Gronwall2007} at magnitudes of $R\gtrsim25$. At $R\lesssim25$, the \citet{Gronwall2007} luminosity function is significantly elevated above our own and approaches the volume densities of LBGs at the same redshift - shown by the filled stars \citep{vanderBurg2010} and pentagons \citep{2009ApJ...692..778R}. Indeed, whilst \citet{2008ApJS..176..301O} report their LAE luminosity function to be $\approx\frac{1}{10}$ of that of the \citet{2009ApJ...692..778R} LBG luminosity function (represented by the dash-dot curve in Fig.~\ref{fig:LF-LAEs-LBG}), \citet{Gronwall2007} find that their LAE $M_{\rm UV}$ luminosity function is equivalent to $\approx\frac{1}{3}$ of the \citet{2009ApJ...692..778R} LBG luminosity function (represented by the dotted curve). Our own LAE $M_{\rm UV}$ luminosity function is consistent with LAEs (of equivalent width $\gtrsim65$~\AA) being $\approx10\times$ less common than LBGs, given magnitudes of $R\lesssim26$ ($M_{\rm UV}\lesssim-19.6$). The difference between the  \citet{Gronwall2007} and our own results would appear to be driven by the differing equivalent width cuts used. This is supported by the distribution function of equivalent widths in LBGs as presented by \citet{shapley03}, whereby there are $\sim3\times$ as many galaxies with equivalent widths in the range $\gtrsim20$~\AA\ (equivalent to the \citealt{Gronwall2007}), than in our range $\gtrsim65$~\AA\ at magnitudes of $R\lesssim25.5$.

We note that as with the number counts in the previous section, we have applied a correction on the assumption that our sample successfully identifies LAEs in 78\% of cases. In the case of this being overly pessimistic, the presented luminosity function could be boosted by up to $\approx25\%$. Within the plotted error bars, this does not change any of the comparisons discussed above.

Again, we also present the GALFORM model results (which are calculated based on our own selection criteria), shown by the dashed curve in Fig.~\ref{fig:LF-LAEs-LBG}. Our observations and the model are consistent within the observational uncertainties at $R\lesssim25$, however the model under-predicts the numbers of galaxies at magnitudes fainter than this limit (contrasting with the narrow band counts in which the model over-predicts the faint end). We note that, when the specific selection constraints of \citet{Gronwall2007} are applied to the GALFORM model, the model under-predicts the continuum luminosity function at both the bright and faint ends, whilst when the same is done with the \citet{2008ApJS..176..301O} selection constraints, the result is the same as seen with our own data. These comparisons between our results and previous observational results and the model will be instructive when analysing the different clustering results.

\begin{figure*}
 	\centering
 	\includegraphics[width=\textwidth]{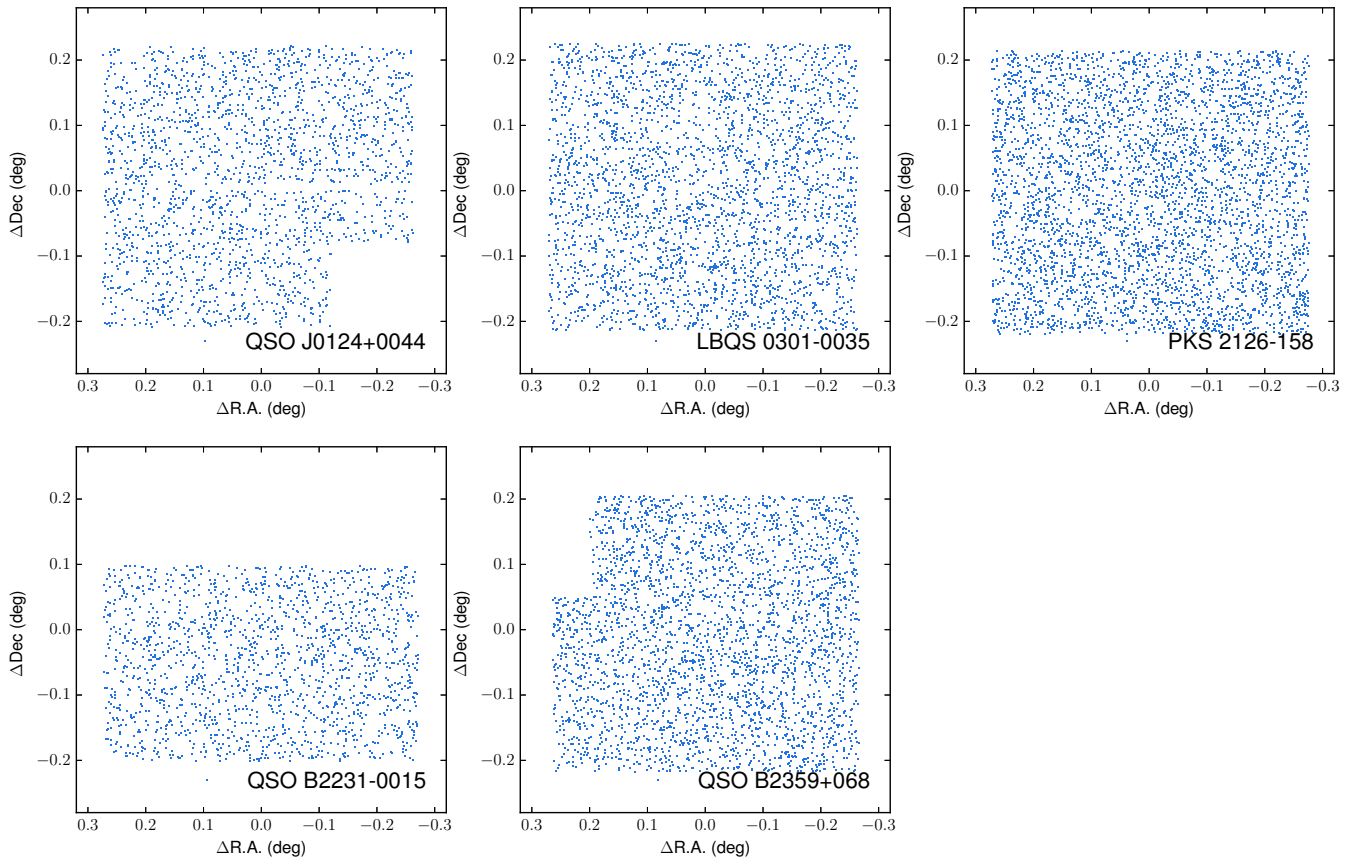}
 	\caption[Random points for LAEs]{The distribution of the random catalogues used in the correlation function analysis, highlighting the masked regions in each of the fields.}
 	\label{fig:Ran}
\end{figure*}

%LAEs are known as star-forming galaxies at $z\approx2-7$ with a faint-ultraviolet (UV) continuum but a prominent Ly$\alpha$ emission line \citep{2008ApJS..176..301O}. A typical star formation rate found in Ly$\alpha$ emitters are $\sim$ 1-10 M$_{\odot}$ yr$^{-1}$ (e.g. \citealt{1998AJ....115.1319C,Hu1998,Gronwall2007,Nilsson2007}). A number of  authors (e.g. \citealt{Ajiki2003,Taniguchi2005,Gronwall2007}) have reported a discrepancy  between the SFR estimated from Ly$\alpha$ luminosity and SFR estimated from UV luminosity. They found larger SFR derived from  UV luminosity than those estimated from Ly$\alpha$ luminosity. However, \citet{Nilsson2007} found SFRs $\sim$ 0.5-6 M$_{\odot}$ yr$^{-1}$ with no discrepancy between continuum and emission line derived SFRs. \citet{Gronwall2007} found the most likely value for the LAE SFR density of the  $z \sim  3.1$ universe  is $8.6 \times 10^{-3}$ $h_{70}$M$_{\odot}$ yr$^{-1}$ Mpc$^{-3}$. They suggested that the true SFR density is probably $\sim$ 3.5 times higher. For a comparison with LBG, the star formation rate density at $z \sim  3.1$ LBG is  $\sim 0.01$ $h_{70}$M$_{\odot}$ yr$^{-1}$ Mpc$^{-3}$ before extinction \citep{Madau1998,steidel99}. \citealt{Gronwall2007}  estimated the  dust-corrected SFR density for LAEs, $\sim 0.03$ $h_{70}$M$_{\odot}$ yr$^{-1}$ Mpc$^{-3}$ which is  $\sim 75\%$  of the LBG value, however, this number is highly uncertain.

Returning to Fig.~\ref{fig:NB-density}, we show the continuum number counts (shifted by $\Delta m=-1.6$, pale red shaded curve) in comparison to the narrow band number counts. In both the narrow band and continuum number counts (and the continuum luminosity function), we see evidence for a turn over at faint magnitudes. Indeed, \citet{2008ApJS..176..301O} for example measure $L^*_{\rm Ly\alpha}\approx5\times10^{42}~{\rm ergs~s}^{-1}$ for their $z\approx3.1$ LAEs, which corresponds to a narrow band magnitude of $m_{\rm NB}\sim24.5$. Our results support the assertion that such LAE samples reaching depths of $m_{\rm NB}\approx25$ are probing the knee of both the NB and continuum luminosity functions. Briefly comparing the narrow-band and continuum number counts, we find that the continuum number counts appear steeper, hinting at an evolution in the EW distribution as a function of galaxy brightness.

\section{LAE Auto-Correlation Function}
\label{sec:clustering}

We now measure the clustering properties of the photometric LAE sample to derive halo masses and other properties.

\subsection{Clustering Estimator}

We estimate the angular auto-correlation function, $w(\theta)$, using the Landy-Szalay estimator \citep{1993ApJ...412...64L}, which is given by:

\begin{equation}
\label{eq-w-theta1}
w(\theta)=\frac{\mathrm{DD}(\theta)-2\mathrm{DR(\theta)}+\mathrm{RR(\theta)}}{\mathrm{RR}(\theta)}
\end{equation}

\noindent where $\mathrm{DD}(\theta)$, $\mathrm{DR(\theta)}$, and $\mathrm{RR(\theta)}$ are the numbers of galaxy-galaxy pairs, galaxy-random pairs and random-random pairs as a function of $\theta$. For each field, we generated uniform random points with the same area as our masked LAE samples, with $20\times$ the number of LAE candidates in each field. Fig.~\ref{fig:Ran} shows the distribution of the random points used in each field for the clustering calculation, showing the extent of the fields and the masking implemented. 

We use two error estimators here: simple Poisson error estimates; and field-to-field error estimates. The Poisson estimate is given by:

\begin{equation}
\sigma_\mathrm{Poi}(\theta)=\frac{1+w(\theta)}{\sqrt{DD(\theta)/2}}
\label{equ:xierr_Poisson}
\end{equation} 
		
The field-to-field error estimate is given by the error on the mean of the measurement across the fields and is calculated using:

\begin{equation}
\sigma_\mathrm{FtF}(\theta)=\sqrt{\frac{1}{N}\frac{1}{N-1} \sum_{i=1}^{N}  
			   [w_{i}(\theta) - \overline{w}(\theta)]^{2}}
\label{equ:xierr_FtF1}
\end{equation}

\noindent where $N$ is the number of fields (i.e. $N = 5$), $w_{i}(\theta)$ is a measurement from the $i$th field and $\overline{w}(\theta)$ is the mean value.

\subsection{Angular Clustering results}
\label{sec:clustering_results}
\begin{figure}
\centering
\includegraphics[width=80.mm]{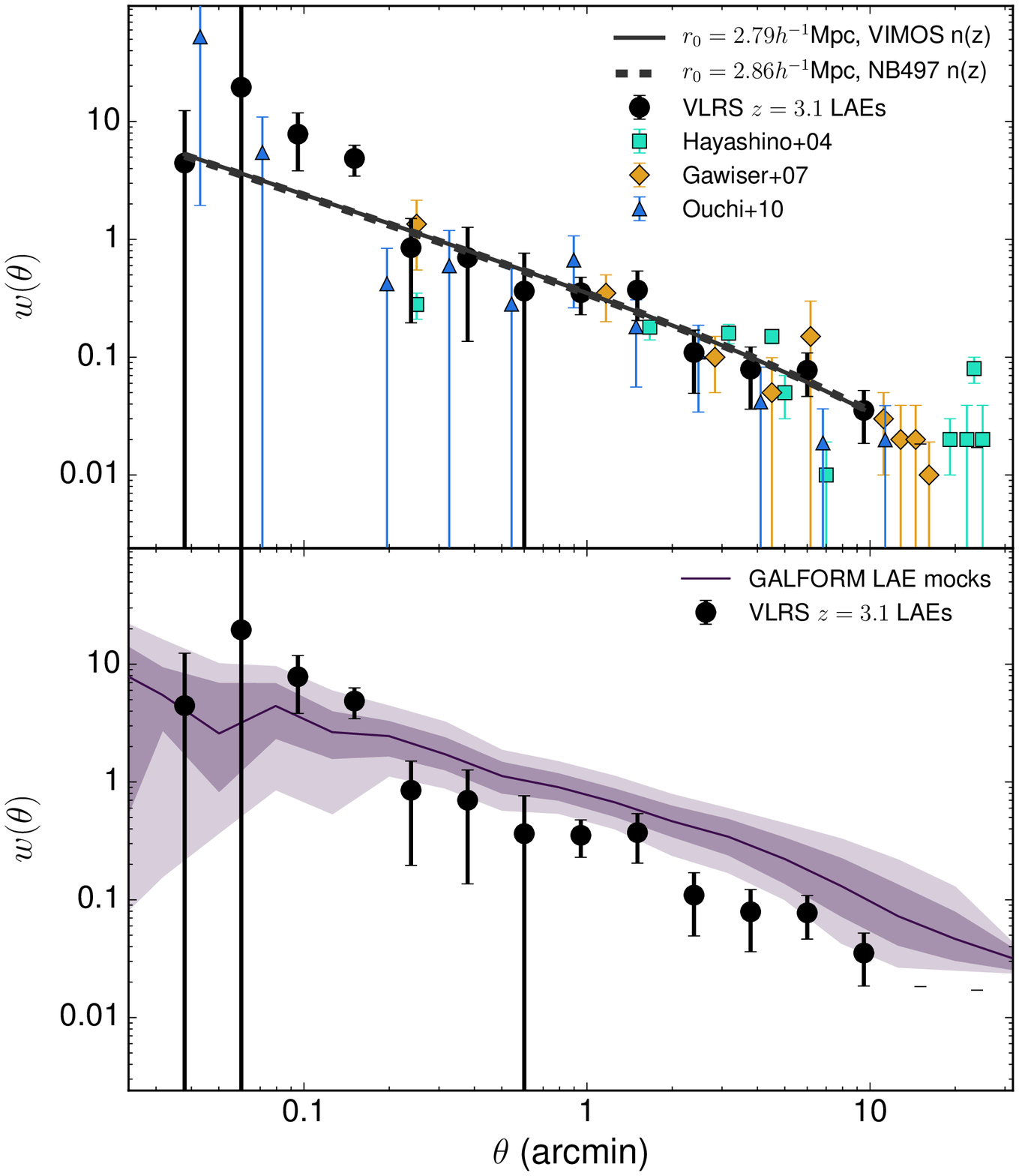}
\caption{\emph{Top panel}: The angular correlation function of the $z=3.1$ LAEs. The filled black circles show the result based on our VLRS LAE sample. Also shown are model fits to the VLRS LAE correlation function based on using the $n(z)$ from our VIMOS spectroscopically observed sample (solid curve) and using an $n(z)$ based on the NB497 filter profile (dashed curve). A number of literature results are also plotted: \citet{2010ApJ...723..869O} is shown by the blue triangles; the \citet{gawiser07} result is shown by the orange diamonds; and the result of \citet{Hayashino2004} is shown by the cyan squares. \emph{Lower panel}: The same VLRS LAE result (filled black circles) alongside the LAE clustering prediction from the GALFORM model. The shaded regions show the 68th and 95th percentile uncertainties based on the standard deviation of 225 1.07~deg$^2$ mock catalogues.}
\label{fig:w_LAELAE}
\end{figure}

Fig.~\ref{fig:w_LAELAE} shows the results of the clustering analysis (filled black circles). We use field to field errors everywhere except at $\theta<1'$, where Poisson estimates are used. The plotted points include an amplitude correction for contamination in the photometrically selected sample, which takes the form:
 
\begin{equation}
 A_{\rm i} = \frac{A_{\rm o}}{(1 - f_c)^2} 
\label{correct}
\end{equation}

\noindent  where $A_{\rm i}$ and $A_{\rm o}$ are the intrinsic and observed correlation amplitude respectively, and $f_c$ is the contamination fraction (i.e. $f_c=0.22$). The plotted points have all been adjusted upwards by a factor of $1/(1-f_c)^2=1.64$. The points are also corrected for the integral constraint, which accounts for the effect of finite field sizes. This is estimated following the method outlined in \citet{2014A&A...568A..24B} and takes a value of ${\cal I}= 0.024$.

% Comparison between our clustering results and others.
% =====================================================
We compare our results in the top panel of Fig.~\ref{fig:w_LAELAE} with several measurements from other authors. Squares show the result from \citet{Hayashino2004} which is obtained from 283 $z=3.1$ LAE candidates with a narrow band magnitude limit of $m_{\rm NB497}<25.8$ and ${\rm EW}_{\rm rest}\geq38$~\AA\ observed in the SSA22a field using the Subaru Telescope. This was a further observation of SSA22a of \citet{Steidel2000} who found 72 LAEs. However, the observed area in \citet{Hayashino2004} is about 10 times larger than in \citet{Steidel2000}. Both \citet{Steidel2000} and \citet{Hayashino2004} concluded that no significant clustering of $z=3.1$ LAEs has been shown inside the SSA22a area. We also compared our results with \citet{gawiser07} (represented by diamonds)  who studied the clustering properties of 162 LAEs at $z = 3.1$   at a narrow-band completeness limit of $m_{5000}\leq25.4$ which were observed in the deep narrow-band MUSYC survey. Their ${\rm EW}_{\rm rest}$ limit is $\approx20~\AA$. \citet{2010ApJ...723..869O} used 356 $z=3.1$ LAEs from  \citet{2008ApJS..176..301O} who carried out a narrowband survey in the 1~deg$^2$ Subaru/XMM-Newton Deep Survery (SXDS) to measure the correlation functionat $m_{\rm NB5003}\leq25.3$ (triangles).

Comparing our results with each measurement, we found that our results agree with \citet{2010ApJ...723..869O} within their error bars. At larger scales, \citet{Hayashino2004} and \citet{gawiser07} show  slightly higher clustering amplitude than ours, but are also in agreement with our results within the error estimates. Given the different equivalent width, and flux limits of each survey, this may be evidence that the clustering amplitude of LAEs is particularly insensitive to the Ly$\alpha$ sample selection. In other words, LAEs trace the density field in the same way regardless of their selection. Care should be taken when comparing the $w(\theta)$ measurements of samples with differing filter widths however. Whilst \citet{Hayashino2004}, \citet{2010ApJ...723..869O}, and our own sample use a filter with FWHM~$\approx75$~\AA, the \citet{gawiser07} LAE sample uses a filter with FWHM~$\approx50$~\AA\. A more reliable comparison can be made when comparing the $r_0$ measurements that follow, which take into account the filter profile. Overall, our results provide the most accurate measurement of the LAE clustering signal to date given our large total area and the use of multiple fields, which minimise any cosmic variance effects.

In the lower panel of Fig.~\ref{fig:w_LAELAE} we compare our results to the predictions of the GALFORM model. The filled circles are the same as in the top panel, whilst the solid curve and shaded regions show the median clustering measured from 225 mocks and the $1\sigma$ and $2\sigma$ uncertainties. Each mock covers a survey area of 1.07~deg$^2$ (i.e. equivalent to our own survey area) and so the uncertainties given by the shaded regions give a further estimate of the uncertainties in our observations (and one that corroborates our estimates based on the data itself).

We find that the model results over-predict the observational measurement at the $\approx2\sigma$ level. Although the semi-analytical model is able to reproduce the narrow-band and continuum luminosity function of LAEs at bright magnitudes, it is worth noting that the clustering results will be dominated by the more numerous faint LAEs, which the model fails to reproduce. It may be that the discrepancy here is connected to the model number count disagreements at faint magnitudes. Taking our results alongside comparable over-predictions of clustering measurements by semi-analytical simulations \citep[e.g.][]{2011MNRAS.413..101G,2013MNRAS.428.1351G,2014A&A...568A..24B}, it is clear that clustering measurements provide an important additional constraint on galaxy formation models. Indeed, resolving these discrepancies at faint magnitudes in the LAE numbers and clustering remains a challenge for the semi-analytical models.

% Add paragraph on how clustering varies between fields here?
% ===========================================================

\subsection{Real-space clustering}
\label{sec:autocorr}
We now parameterise the LAE clustering in terms of the real-space clustering length as measured via the angular two-point correlation function $w(\theta)$, which is effectively a weighted projection of the spatial two-point correlation function $\xi(r)$. It is common to transform from the spatial to the angular correlation function using Limber's approximation \citep{1953ApJ...117..134L}, however this requires that the depth of the galaxy survey, $\pi$ (i.e. distance probed along the line of sight), is much greater than the on sky maximum separation, $\sigma$ \citep[e.g.][]{1953ApJ...117..134L,2007A&A...473..711S}. For our survey volume, this is not the case - the depth of our survey is $\pi\approx60~h^{-1}$Mpc and the width is $\sigma\approx40~h^{-1}$Mpc. We therefore use the full analytical form in transforming from the real-space clustering form to the projected angular clustering.

As the underlying model for the real-space clustering, we assume a power-law with a slope of $\gamma=1.8$ \citep[e.g.][]{1980lssu.book.....P}:

\begin{equation}
\xi(r) = \left(\frac{r}{r_0}\right)^{-1.8}
\end{equation}

\noindent where $r$ is the real-space separation between two galaxies and $r_0$ is the clustering length parameter.

The spatial correlation function, $\xi(r)$ can be related to $w(\theta)$ by \citep[e.g.][]{Phillipps1978}:

% 'exact analytical solution'.
 \begin{equation}
 w(\theta)=\frac{\int_{0}^{\infty}{dz_{1}f(z_{1})}\int_{0}^{\infty}{dz_{2}f(z_2)
 \xi(r)}}{\left[\int_{0}^{\infty}dzf(z)\right]^2}
 \label{mylim}
 \end{equation}

\noindent where $f(z)$ is the radial distribution of sources which is given by:

\begin{equation}
f(z) \equiv \chi^2(z)\frac{d\chi(z)}{dz}n_c(z)\phi(z)
\end{equation}

\noindent where $\phi(z)$ is the selection function of the sample, $n_{c}(z)$ is the comoving number density of the sources, $\chi$ is the radial comoving distance, and $r=r(\theta,z_1,z_2)$ is a comoving separation between two points at $z_1$ and $z_2$.

\begin{equation}
r \equiv \sqrt{\chi^2(z_1)+\chi^2(z_2)-2\chi(z_1)\chi(z_2)\cos \theta}
\end{equation}

We assume that over the redshift range probed, the actual number density of LAEs, $n_c(z)$, is constant, whilst we have calculated solutions with each of the filter response curve and the redshift distribution of the VIMOS detected sources as the selection function, $\phi(z)$. We performed a minimised $\chi^2$ fit to determine the clustering length, $r_0$, given our $w(\theta)$ clustering measurement and using Eq.~\ref{mylim} to transform from the real-space power law correlation function, $\xi(r)$, and the angular correlation function, $w(\theta)$. With the $n(z)$ of the spectroscopically confirmed LAEs and a fixed slope of $\gamma=1.8$, we derived a result of $r_{0}=2.79\pm0.34\hmpc$. The resulting $w(\theta)$ model is shown by the solid curve in Fig.~\ref{fig:w_LAELAE}. With the selection function given by the filter transmission curve, we found a best fitting value of $r_{0}=2.86\pm0.33~h^{-1}{\rm Mpc}$ - shown by the dashed curve in Fig.~\ref{fig:w_LAELAE}. Given that these two estimates of $r_0$ are consistent with each other within the estimated uncertainties, it is clear that the selection function is not significant for the result. We note that the data-points appear to rise above the fitted model at small separations ($\theta\lesssim0.15''$), potentially indicating the beginning of the one halo term. Indeed the scale at which this turn-up in the correlation function is seen is comparable to that reported for $z\approx3$ LBGs by \citet{2009A&A...498..725H}.

Such a low clustering measurement is consistent with previous measurements of the clustering of $z\approx3$ LAEs. Those shown in Fig.~\ref{fig:w_LAELAE} have reported clustering lengths of $r_{0}=2.5^{+0.6}_{-0.7}\hmpc$ \citep{gawiser07} and $r_{0}= 1.70^{+0.39}_{-0.46}\hmpc$ \citep{2010ApJ...723..869O}, whilst \citet{Hayashino2004} did not present a measurement of the clustering length. As discussed, the \citet{2010ApJ...723..869O} sample is based on a comparable EW cut as our own sample and so presents the most like-for-like comparison. Compared to their "best" estimate of the clustering length, we measure a moderately higher $r_0$, at the $\approx2-3\sigma$ level. Both measurements are subject to uncertainties in contamination and they report a "maximum" clustering length estimate of $r_0=1.99^{+0.45}_{-0.55}$. Similarly, our own measurement may be 'over-corrected' for contamination based on the spectroscopic non-detections, i.e. if the non-detections were in fact LAE too faint to be detected in the VIMOS observations. If this were the case, and we assume no contamination of our sample, then we would measure a clustering length of $r_0=2.18~\hmpc\pm0.26$, somewhat more consistent with the previous measurement. 

The LAE clustering is low compared to both the normal star-forming and passive galaxy populations at $z\sim2-3$. Indeed, at $z\sim3$, the LBG population of star-forming galaxies is found to have clustering lengths of $r_0\sim4\hmpc$ \citep[e.g.][]{2005ApJ...619..697A,2011MNRAS.414....2B,2013MNRAS.430..425B}. Given literature results suggesting a strong link between galaxy stellar mass and median halo mass \citep[e.g.][]{2014A&A...568A..24B,2015MNRAS.449..901M}, such a low clustering length is strongly suggestive of LAEs being low-mass galaxies residing in low-mass dark matter halos.

\subsection{Dark Matter Halo Masses}

We now explicitly estimate median halo masses from our clustering results. Following \citet{2014A&A...568A..24B}, we calculated the mean masses of dark matter halos within the galaxy samples by using the clustering results. We used the formalism developed by \citet{mowhite96} which provides a relationship between the halo-bias to the mean halo mass via \citet{2001MNRAS.323....1S}. Based on the measured clustering length derived using the filter transmission profile, we find a clustering bias of $b=2.13\pm0.22$ and halo mass $M_{\rm DM}=10^{11.0\pm0.3}~h^{-1}$M$_{\odot}$.

\citet{gawiser07} reported a bias factor of $b=1.7\pm0.4$ and a median dark matter halo masses of $M_{\rm DM}=10^{10.9\pm0.9}$~M$_{\odot}$, whilst \citet{2010ApJ...723..869O} reported a bias value of $b$ = 1.7$\pm$0.8 and a halo mass $M_{\rm DM}\approx6.7^{+42.0}_{-6.7}\times10^{10}$~M$_{\odot}$. We note that, although the cosmologies used in these papers and our own differ slightly, this only affects the halo mass values by a factor of $\approx10^{0.1}$. \citet{2010ApJ...723..869O} concluded that the average dark halo mass of LAEs is $\sim 10^{11\pm1}$~M$_{\odot}$ at $z=2-7$ and our own measurement is more consistent with this assertion than their own results at $z=3.1$. As for the GALFORM model clustering result shown in Fig.~\ref{fig:w_LAELAE}, this corresponds to LAEs occupying dark matter halos with a median mass of $M_{\rm DM}\approx10^{11.3\pm0.3}~h^{-1}$M$_{\odot}$, approximately $1\sigma$ higher than the measurement based on our observations.

The median halo mass for LBGs is estimated to be $\sim10^{12}$~M$_{\odot}$, about one order of magnitude larger than that of LAEs \citep{Hamana2004,Ouchi2004b,Ouchi2005b,2006ApJ...642...63L,Lee2009,McLure2010,2009A&A...498..725H}. Indeed, from our own LBG measurements, we estimate a median LBG halo mass at $z\sim3$ of $M_{\rm DM}=10^{11.6\pm0.2}\hmodot$ with $b=2.37\pm0.21$ \citep{2013MNRAS.430..425B}, larger than the LAE measurements at the $\approx2\sigma$ level. 

\subsection{LAE-LBG cross-correlation}

We now turn to the LAE-LBG cross-correlation, combining the LAE candidates selected in this paper with the spectroscopically confirmed sample of LBGs presented by \citet{2013MNRAS.430..425B}. Combining the cross-correlation of the two samples with their respective auto-correlation samples, we may infer the relationship between the dark matter density distributions that they each inhabit \citep[see for example][]{2006MNRAS.367.1251R,2014MNRAS.437.2017T}.

From the Cauchy-Schwartz inequality \citep[e.g.][]{cauchy1821cours,mitrinovic2013}, \citet{adelberger03} derived the following inequality for the auto- and cross-correlation functions of the two populations:

\begin{equation}
\xi_{\rm ab}^2 \leq \xi_{\rm aa}\xi_{\rm bb}
\label{eq:inequality}
\end{equation}

\noindent where $\xi_{\rm ab}$ is the cross correlation and $\xi_{\rm aa}$ and $\xi_{\rm bb}$ are the respective auto-correlation functions. As discussed (in reference to the galaxy-H{\sc i} cross-correlation) in \citet{2014MNRAS.437.2017T}, if the equality holds, the two samples are deemed to trace the same underlying mass distribution, and the relative biases can be used to infer the ratio between the dark matter halo populations that the two populations occupy. 

We calculate the cross-correlation between the photometric LAE sample and a subset of the LBGs matching the criteria $|z_{\rm LBG}-3.08|\leq0.08$ (giving $\approx80$ LBGs). Although this leaves a relatively small number of LBGs, it optimises the cross-correlation signal by excluding poorly correlated pairs at large separations along the line of sight. For the calculation, we use the Landy-Szalay estimator:

\begin{equation}
\small w(\theta) = \frac{\left<D_aD_b(\theta)\right>-\left<D_aR_b(\theta)\right>-\left<R_aD_b(\theta)\right>+\left<R_aR_b(\theta)\right>}{\left<R_aR_b(\theta)\right>}
\end{equation}

\noindent where $D_aD_b$ is the LBG-LAE pair count, $D_aR_b$ is the LBG-random LAE pair count, $D_aR_b$ is the random LBG-LAE pair count and $R_aR_b$ is the random LBG-random LAE pair count (all normalised by the number of randoms per galaxy). We use random catalogues with $20\times$ the number of randoms as galaxies in each case.

\begin{figure}
\centering
\includegraphics[width=80.mm]{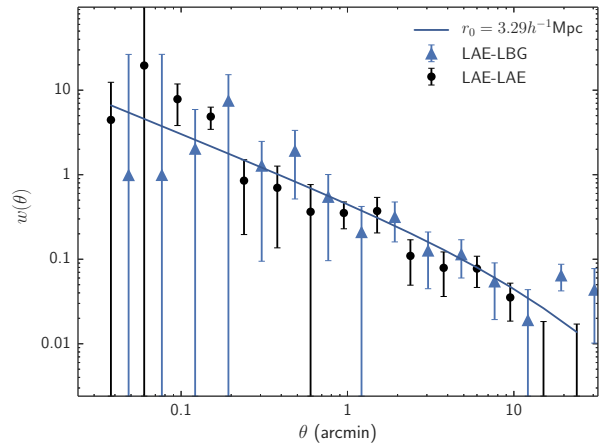}
\caption{The LAE-LBG angular cross-correlation function calculated using the VLRS LAE and LBG samples (triangles). A fit to the data using a single power-law form (with a correlation length of $r_0=3.29\pm0.57$) is shown by the solid curve. For comparison, the LAE auto-correlation is reproduced (circles).}
\label{fig:xw_LBGLAE}
\end{figure}

The results are shown by the triangles in Fig.~\ref{fig:xw_LBGLAE}, alongside the LAE auto-correlation result (filled circles). We fit the cross-correlation result based on a single power-law form in real-space, i.e. $\xi(r)$, transformed to $w(\theta)$ based on the formalism outlined in Sec.~\ref{sec:autocorr} (but now incorporating the two separate redshift distributions rather than just that of the LAE). The best-fitting result is shown by the solid curve in Fig.~\ref{fig:xw_LBGLAE}, corresponding to a correlation length of $r_0=3.29\pm0.57~h^{-1}$Mpc (we do not fit for the slope, but take a fixed value of $\gamma=-1.8$). This compares to a LAE auto-correlation clustering length of $r_0=2.86\pm0.33~h^{-1}$Mpc and a LBG auto-correlation clustering length of $r_0=3.83\pm0.24~h^{-1}$Mpc \citep[with a slope of $\gamma=-1.6$,][]{2013MNRAS.430..425B}.

Evaluating Eq.~\ref{eq:inequality} using the clustering amplitudes corresponding to $r=1~h^{-1}$Mpc (i.e. $r_0^\gamma$), we find $\xi_{\rm ab}^2/\xi_{\rm aa}\xi_{\rm bb}=1.28\pm0.46$, i.e. $\approx1$. As discussed by \citet{adelberger05,2014MNRAS.437.2017T}, this unity in the relation does indeed imply that the two photometric selections give samples of galaxies that consistently trace the same underlying dark matter distribution and we can therefore use this result to provide a second constraint on the typical halo mass that our LAE sample occupies. Similar to the auto-correlation case, the correlation function can now be related to the underlying dark matter correlation function by:

\begin{equation}
\xi_{\rm LAE-LBG}=b_{\rm LAE}b_{\rm LBG}\xi_{\rm DM}
\end{equation}

\noindent where $\xi_{\rm LAE-LBG}$ is the LAE-LBG cross-correlation, $b_{\rm LAE}$ is the LAE bias, $b_{\rm LBG}$ is the LBG bias, and $\xi_{\rm DM}$ is the underlying dark matter clustering as in the previous section. Based on this and our cross-clustering results, we calculate a further measure of the LAE bias of $b_{\rm LAE}=2.24\pm0.37$. Using this to estimate a halo mass as in the previous section, we find a value of $M_{\rm DM}=10^{11.1\pm0.4}~h^{-1}$M$_{\odot}$. The cross-correlation results corroborate the auto-correlation analysis with both dark matter halo estimates being consistent.

\section{Discussion}
\label{sec:discussion}

Our results, based on the largest $z\sim3$ narrow band LAE sample yet surveyed and the first to combine multiple independent large fields, builds on the results presented by others \citep[e.g.][]{Gronwall2007,gawiser07,2008ApJS..176..301O,2010ApJ...723..869O} to add to the overall picture of the nature of LAE galaxies at this epoch.

At first sight, LAEs at $z\approx3$ seem to be a high density population with a sky density of $\approx0.1$arcmin$^{-2}$ in a redshift range of width only $\Delta z\approx 0.05$ to the various limits used in our survey. Taking the sky density of LBGs to be $\approx1.8$ arcmin$^{-2}$  in the $2.5<z<3.5$ redshift range \citep{steidel03}, this corresponds to a roughly similar sky density. On the other hand, comparing galaxy numbers at fixed continuum luminosity  \citep[e.g.][Fig.~\ref{fig:LF-LAEs-LBG} in this paper]{2008ApJS..176..301O} we confirm that LAE are a relatively rare component of the galaxy population with abundances $\approx10\times$ lower than that of LBGs.

Given their relative rareness, one might expect that LAEs would occupy rare high-mass dark matter halos based on a simple abundance matching approach. Our clustering results show a somewhat higher clustering signature than the previous best study of LAEs at this redshift \citep{2010ApJ...723..869O}, however the clustering results still clearly show that the LAEs are relatively poorly clustered compared to average star-forming galaxies at $z\approx3$ \citep[e.g.][]{2005ApJ...619..697A,2013MNRAS.430..425B,2015A&A...583A.128D}. Indeed our measurement of LAE clustering implies a median halo mass of $M_{\rm DM}\approx10^{11}~h^{-1}{\rm M}_{\odot}$, for which the number densities are predicted to be $\sim0.008~h^3{\rm Mpc}^{-3}$ (integrating the mass function over the mass range $10^{11\pm0.1}~\hmpc$) in a $\Lambda$CDM Universe \citep[][]{2001MNRAS.321..372J,2013A&C.....3...23M}. It therefore follows (given our measured number density of LAEs of $\approx2\times10^{-3}~h^{3}{\rm Mpc}^{-3}$) that LAEs possess a relatively low occupation number (i.e. number of galaxies per halo) of $<<1$ and trace the dark matter distribution relatively sparsely (corroborating the results of \citealt{2010ApJ...723..869O}). It is common to link a galaxy sample's properties as far as possible to its host halo properties, but it is apparent that the galaxy properties that govern LAE selection (i.e. Ly$\alpha$ equivalent width) are unlikely to be well correlated with the halo mass. Indeed, our results support the hypothesis that LAEs presence within dark matter halos is stochastic in nature \citep{2010PASJ...62.1455N,2010ApJ...723..869O}.

Our measured halo mass of $M_{\rm DM}=10^{11.0\pm0.3}~h^{-1}{\rm M}_{\odot}$ is consistent with estimated LAE halo masses across a wide redshift range of $2<z<7$ (i.e. the full range of redshifts probed by available LAE samples). Indeed, our measurement brings a tighter agreement with available measurements at other redshifts than the previous best measurement at $z\approx3$ \citep[i.e.][]{2010ApJ...723..869O}. As such, this enhances the assertion made by \citet{2010ApJ...723..869O} that LAEs exist as a stage in the evolution of at least some galaxies upon the host dark halo reaching a mass of $M_{\rm DM}\approx10^{11}~h^{-1}{\rm M}_{\odot}$.

The LAE sample as selected using the NB497 filter lies within the
redshift range of LBGs selected based on their $U-B$ (or $U-G$) colours.
It is worth considering then how the galaxies identified by these two
selections relate to each other. LBGs by definition are galaxies that
possess a Lyman Break, however this is not intrinsic to the galaxies
themselves and so LAEs are also LBGs in the sense that they will have a
Lyman break of some form, although it still may be too weak to satisfy
the LBG cut \citep{2014MNRAS.441..837C}. Nevertheless, since the LAE
number density at fixed continuum luminosity is $\approx10\times$ lower
than that of the LBGs, the LAE selection appears simply as an extreme subset of the LBG selected population. Indeed 40\% of the LBG population has been shown to exhibit Ly$\alpha$ in emission, with $\approx25\%$ of LBGs having equivalent widths of $\gtrsim20$~\AA\
\citep{shapley03} and a tail in Ly$\alpha$ equivalent width up to
$\sim200$~\AA. Clearly, at least some LAEs would fall into an LBG
selected sample and the numbers seem to approximately tally given that
\citet{Gronwall2007} reported number densities of $\gtrsim20$~\AA\
equivalent width LAEs were approximately a third of the number densities
of LBGs. Indeed, Fig.~8 of \citet{shapley03} shows that $\approx12\%$ of
their LBG sample have equivalent widths of $\gtrsim60$~\AA, which ties
in very well with the number densities we find for our sample (as also noted by \citealt{2008ApJS..176..301O} for their own $z\approx3.1$ ${\rm EW}\gtrsim60$~\AA\ LAE sample).

%In Fig.~\ref{fig:lae-ubr} we show the $U-B$ versus $B-R$ colour plot for our LAE sample (filled circles and triangles) compared to the VLRS LBG selection in these fields (to the upper left of the solid line boundary) and the general $R<25$ galaxy population (shaded 2d histogram). Given that LAEs are selected to have faint continua, the uncertainties on the LAE colours are relatively large, whilst many have no $U$ detection at all (shown as lower limits on the $U-B$ colour by the triangle points). Of the LAEs shown, 60\% fall within the LBG selection region or present lower limits consistent with the LBG selection. Of the 40\%, that do not, only 2 galaxies lie outside the LBG selection region with any significance given the estimated photometric uncertainties. Thus the true percentage lies in the range $\approx60-90\%$.
%
%\begin{figure}
%\centering
%\includegraphics[width=\columnwidth]{LAE_UBR_chart.eps}
%	\caption{The $U-B$ versus $B-R$ colour diagram for the general $R<25$ galaxy population (shaded 2d histogram) and for the spectroscopically identified $z\approx3.1$ LAE (filled circles and triangles). Galaxies selected as LBGs lie in the region to the top left of the solid line boundary. Many of the points have no $U$-band detection and so are simply plotted as lower limits, whilst $\approx\frac{1}{2}$ of our sample have either no $B$ or no $R$ band detection and so are not plotted.} 
%\label{fig:lae-ubr}
%\end{figure}

\begin{figure}
\centering
\includegraphics[width=\columnwidth]{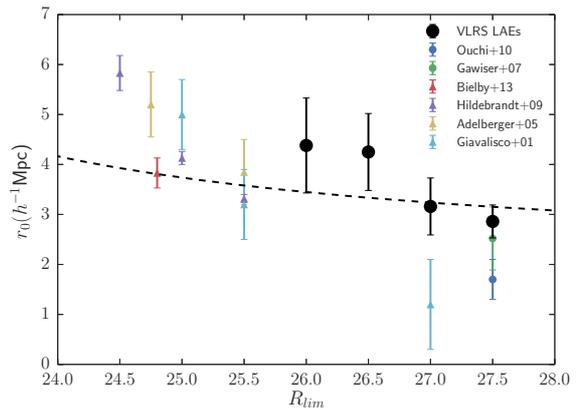}
	\caption{We show the clustering results as a function of limiting continuum magnitude for our LAE sample (large circles) and a range of $z\approx3$ LAE (circles) and LBG (triangles) samples. The dashed curve shows the predicted luminosity dependence based on a 1:1 relationship between luminosity and halo mass.} 
\label{fig:lumVr0}
\end{figure}

Our clustering analysis corroborates this picture of the LAE selected sample being an extreme subsample of the LBG population, whereby they show lower clustering due to our LAE sample probing a low-luminosity subset of the LBG population (at least compared to the mostly $R\lesssim25$ LBG samples from which clustering measurements have been made). This is consistent with the UV luminosity dependence of the observed LBG clustering, whereby LBG samples selected with fainter UV luminosities are observed to have lower clustering signals \citep[e.g.][]{2001ApJ...550..177G,2006ApJ...642...63L,Lee2009,2008ApJ...679..269Y}. In Fig.~\ref{fig:lumVr0}, we show the clustering results taken from a number of LBG (triangles) and LAE (circles) surveys including our own LBG clustering measurement and measurements of the clustering of our LAE sample when limited by a range of continuum magnitudes (filled black circles). The literature $z\approx3$ LAE results shown are from \citet{gawiser07,2010ApJ...723..869O}, and the $z\sim3$ LBG samples are from \citet{2001ApJ...550..177G,2005ApJ...619..697A,2009A&A...498..725H,2013MNRAS.430..425B}. The dashed curve shows the predicted $r_0$-magnitude dependence assuming a 1:1 ratio between UV continuum luminosity and halo mass, and using the \citet{mowhite96} formalism to derive clustering lengths.

We find a tentative continuum luminosity dependence in our LAE results, albeit only at the $\approx1\sigma$ level. Taking our own LBG clustering results (which include the 5 fields used for the LAE study in this paper), our results are in broad agreement since our LBG limit is $R_{AB}\approx24.8$ and our lowest LAE continuum limit in Fig.~\ref{fig:lumVr0} is $R_{AB}\approx 27$ or a factor of $\approx8$ in UV luminosity. Assuming that luminosity is approximately proportional to mass, given the above halo model this represents a factor of $\approx 1.5$ in bias. This compares to the observed factor of $1.35\pm0.2$ given the measured $r_0$ for LBG and LAE that we have measured above. The LAE clustering length of $r_0=2.86\pm0.33~h^{-1}$Mpc is consistent with the decreasing trend of LBG clustering length measured by \citet{2009A&A...498..725H}.  The VLRS LAE and LBG samples show an $R_{\rm lim}$-$r_0$ dependence consistent with this simple model. At brighter magnitude cuts however, a steeper dependence is exhibited by the results of \citet{2005ApJ...619..697A} and \citet{2009A&A...498..725H}.

Larger LAE datasets are required to test if at the $R_{AB}\approx25$ LBG limit, LAEs and LBGs have the same clustering amplitude, but at this point it seems that there is consistency with a model where LAEs can be treated as a subset of LBGs and where halo mass is proportional to UV luminosity in this combined population. The LBGs have a spatial number density that is consistent with their halo masses implied from clustering but the LAE number density is too low, implying that they can only randomly sample their potential halo hosts. From the cross-correlation results and the equality of the ratio between the cross-correlation and the respective auto-correlations, it is also evident that although the fainter LAEs only random sample  smaller mass  halos, both the LBG and LAE populations  trace the underlying dark matter distribution in a linear way but with significantly different bias.

% In addition, the consistency shown
%between the cross-correlation and auto-correlation analyses provides
%credence to our estimates of the success of our LAE selection.

%The LAE selection presents a very incomplete sampling of the typical
%dark matter halos that host them however, due to not all galaxies in
%equivalent mass halos exhibiting such strong Ly$\alpha$ emission.
%Speculating on the reason of this low occupation number, one possibility
%is that Ly$\alpha$ emission is tracing star formation episodes within
%$\approx$10 Myrs, whereas the UV continuum in LBGs traces star formation
%over a few hundred Myrs. Given the added complexity of the radiative
%transfer of Ly$\alpha$ photons, this could explain why a galaxy emitting
%strong observable Ly$\alpha$ emission can be thought of as difficult. If
%there are environmental effects, such as star formation quenching in the
%vicinity of a QSO, then this would also impact the visibility of
%Ly$\alpha$ flux over the UV continuum.

\section{Conclusions}
\label{sec:conclusions}

In this paper, we have presented NB497 band photometric observations within 5 of our VLRS fields, taken using Subaru Suprime-Cam. Combining these narrow band images with $B$, $V$ and $R$ deep imaging, we have made a photometric selection of $z\approx3.1$ LAEs with an equivalent width limit of $\approx65$~\AA. We have also made spectroscopic follow-up observations of LAE candidates in 3 of the 5 fields, using VLT VIMOS. Two factors mean that this survey improves on previous work on the study of LAEs at comparable epochs: firstly our choice of 5 independent survey fields minimises the potential effects of cosmic variance on our results compared to previous work; and secondly that our fields contain significant $z\approx3$ LBG data has allowed us to perform a first cross-correlation analysis of the two populations.

The major findings of our study are summarised as follows:

1. Our final LAE selection, based on the combination of $B-m_{\rm NB497}$ with either $V-m_{\rm NB497}$ or $R-m_{\rm NB497}$, gives a success rate of $78\pm18\%$ based on our spectroscopic observations. The equivalent width limit of our final selection is $\approx65$~\AA, whilst the flux limit is $\approx2\times10^{-17}~{\rm erg}~{\rm cm}^2~{\rm s}^{-1}$ (equivalent to $L\approx10^{42}~{\rm erg}~{\rm s}^{-1}$). A catalogue of the photometric LAEs from the 5 VLRS fields presented here is available at the Strasbourg Astronomical Data Center (CDS, \url{http://cds.u-strasbg.fr/}).

2. The spectroscopic observations produced 35 confirmed LAEs at $z\approx3.1$, with 23 of these falling within our final optimised selection criteria. We found no strong evidence for any contamination from low redshift O{\sc ii} emitters (although $\lesssim5\%$ of the detections remain ambiguous given the observational constraints). The only significant `contamination' of the selection is from objects with no discernible emission in the observed spectra, pointing to either low Ly$\alpha$ flux objects and/or interlopers (both likely entering the selection due to photometric uncertainties).

3. We have calculated the NB497 number counts for our LAE sample (corrected for contamination) and find it to be consistent within the uncertainties with the previous observations of \citet[][equivalent width limit $\approx20$~\AA]{Gronwall2007} and \citet[][equivalent width limit $\approx65$~\AA]{2008ApJS..176..301O}. Our selection is shown to probe the knee of the LAE Ly$\alpha$ luminosity function based on an observed turnover in the narrow band number counts at $m_{\rm NB}\sim24.5$.

4. We derived the $R$-band/rest-frame UV continuum luminosity functions
of our sample of LAEs, again corrected for contamination. Our luminosity
function lies at a level $\approx10\times$ lower than that of the
$z\approx3$ LBG luminosity function, consistent with the LAE population
being a small subset of the LBG population. At $R\lesssim26$ ($M_{\rm
UV}\lesssim-19.6$) our results are consistent with the fraction of
strong emitters being constant or marginally increasing towards fainter
continuum luminosities, consistent with the results of
\citet{2008ApJS..176..301O}. Over a comparable magnitude range, we find
our UV continuum luminosity function is significantly lower (by a factor
of $\gtrsim3$) than that measured by \citet{Gronwall2007}, however this
appears consistent with their lower equivalent width threshold of
$\approx20$~\AA.

5. We have measured the angular correlation function for our $z\sim3.1$
LAE photometric sample in our 5 fields. Our results (corrected for
contamination) are significantly lower than the observed clustering of
LBGs at the same redshift. We measure a clustering length of
$r_0=2.86\pm0.33~h^{-1}{\rm Mpc}$ (assuming a slope of $\gamma=1.8$),
which corresponds to a clustering bias of $b=2.13\pm0.22$ and a median
halo mass of $10^{11.0\pm0.3}~h^{-1}{\rm M}_\odot$. This indicates that
LAEs reside in low-mass dark matter halos, but given their number
densities they have a low occupation number - i.e. only a small fraction
of such halos actually host an LAE.

6. We measure the LBG-LAE cross-correlation function for the first time and find the
results to be in agreement with the auto-correlation analysis, providing
a useful consistency check on our results. From the cross-correlation
analysis, we find a bias for the LAE population of $b=2.24\pm0.37$ and a
typical dark matter halo mass of $M_{\rm DM}=10^{11.1\pm0.4}~h^{-1}$M$_{\odot}$.

7. Thus our results for LAE number counts, clustering and cross-clustering are in agreement with a view that LAE, as selected here, are a low luminosity subset of LBGs that inhabit low mass dark matter haloes with a small occupancy. Comparison to current semi-analytical models show good agreement in the narrow band counts, but poorer agreement in the LAE clustering properties. It will be interesting to see how future changes to semi-analytical models can better match the observational results.

\section*{Acknowledgements}

RMB \& TS acknowledge the funding of their work through the UK STFC research grant ST/L00075X/1. PT acknowledges financial support from the Royal Thai Government. Leopoldo Infante and Jorge Gonz\'alez-L\'opez obtained partial support from CATA, Conicyt Basal program. The plots in this publication were primarily produced using Matplotlib \citep[][\url{http://www.matplotlib.org}]{2007CSE.....9...90H}. We finally thank Prof. Tomoki Hayashino for providing access to the NB497 filter for our observations.

\bibliographystyle{mnras}
\bibliography{$HOME/Documents/lib/rmb}

\appendix

\section{VLT VIMOS 1D and 2D spectra of confirmed LAEs}
% Spectra plots created using exthires1d-new.pro

The figures show the observed-frame spectra taken using VLT VIMOS with the HR\_Blue grism as part of our spectroscopic follow-up observations. All the spectra shown have been identified as LAEs with significant emission within the NB497 filter passband.

	\begin{figure*}
	\centering
	\includegraphics[width=\textwidth]{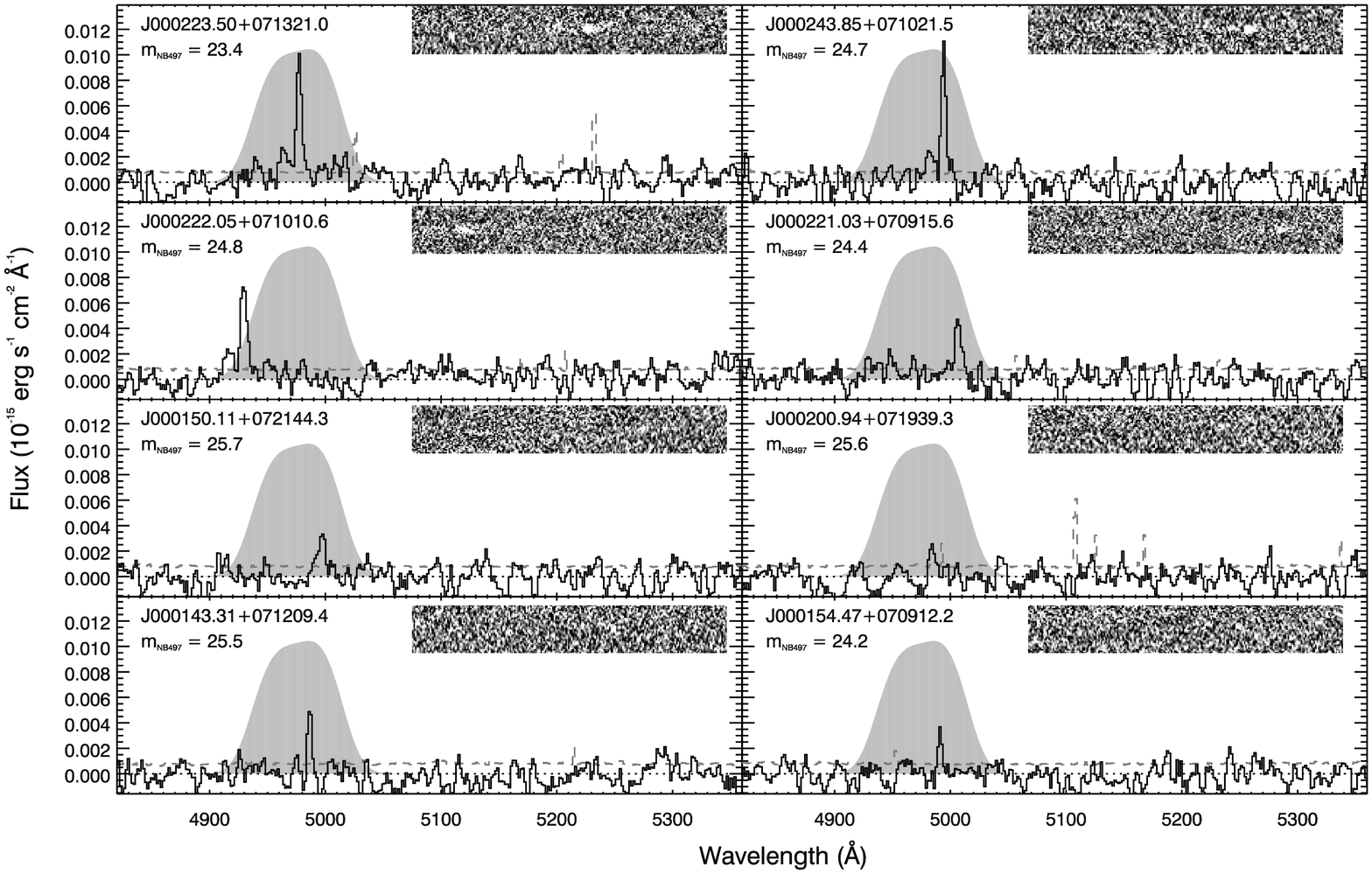}
	\caption[Q2359 Vimos spectra.]{VIMOS HR\_red spectra (black solid curve) for confirmed LAEs in the QSO B2359$+$0653 field. The grey filled region shows the transmission profile of the NB497 filter. In addition, the insets show the corresponding 2D spectra showing only the wavelength region covered by the filter. The dashed lines show the noise estimated from the standard deviation of the background signal across the slit.} 
	\label{fig:Q2359-vimos-spectra}
	\end{figure*}

	\begin{figure*}
	\centering
	\includegraphics[width=\textwidth]{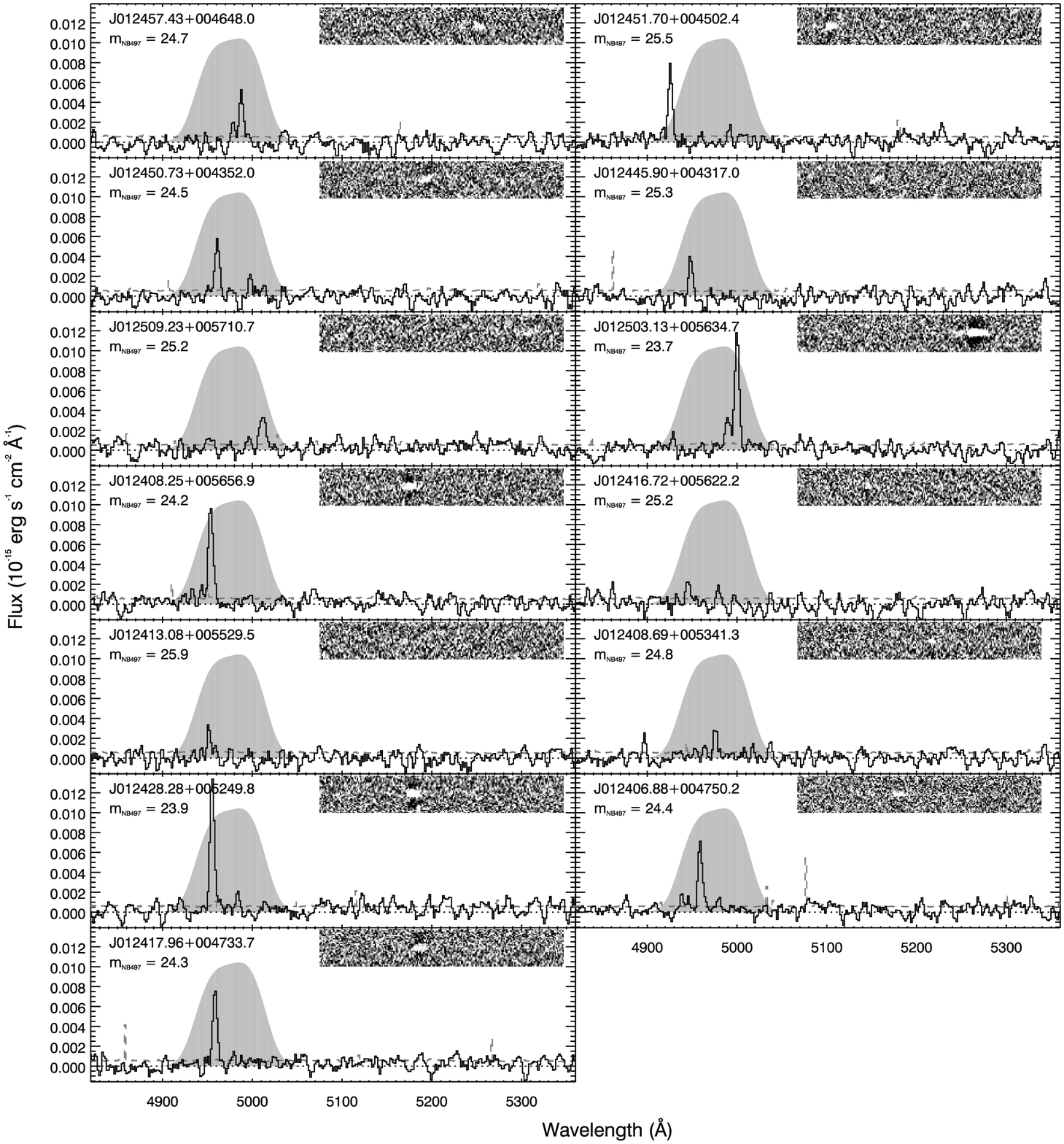}
	\caption[J0124 Vimos spectra.]{As for Fig.~\ref{fig:Q2359-vimos-spectra}, but for QSO J0124$+$0044 field.} 
	\label{fig:J0124-vimos-spectra}
	\end{figure*}

	\begin{figure*}
	\centering
	\includegraphics[width=\textwidth]{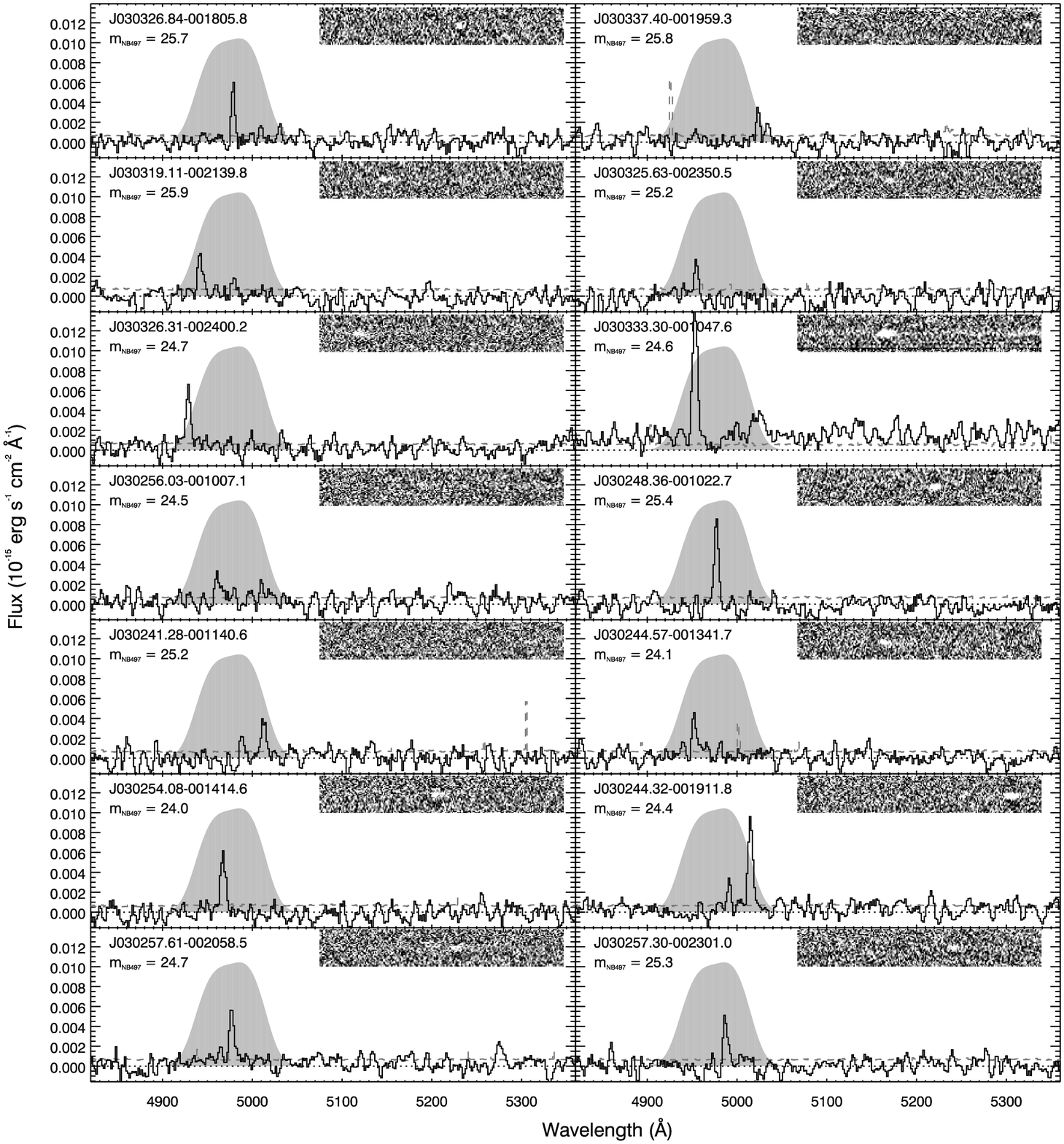}
	\caption[Q0302 Vimos spectra.]{As for Fig.~\ref{fig:Q2359-vimos-spectra}, but for LBQS Q0301$-$0035 field.} 
	\label{fig:Q0302-vimos-spectra}
	\label{lastpage}
	\end{figure*}

\end{document}